\documentclass[sigconf]{acmart}
\usepackage[normalem]{ulem} 
\setcopyright{none}
\renewcommand\footnotetextcopyrightpermission[1]{}

\newcommand{\hide}[1]{}
\newtoggle{showred}
\togglefalse{showred}

\newtoggle{showcrossedtext}
\togglefalse{showcrossedtext}

\newcommand{\myred}[1]{\iftoggle{showred}{\textcolor{ACMRed}{#1}}{#1}}
\newcommand\redsout{\bgroup\markoverwith{\textcolor{red}{\rule[0.5ex]{2pt}{0.4pt}}}\ULon}

\hide{
\usepackage{fancyhdr}
\fancypagestyle{plain}{%
   \fancyhf{} %
   \fancyfoot[L]{ACM SIGCOMM Computer Communication Review}%
   \fancyfoot[R]{Volume 48 Issue 1, January 2018}%
}
\pagestyle{plain}

\fancypagestyle{firstpagestyle}{%
   \fancyhf{} %
   \fancyfoot[L]{ACM SIGCOMM Computer Communication Review}%
   \fancyfoot[R]{Volume 48 Issue 1, January 2018}%
}
}

\usepackage{balance}

\PassOptionsToPackage{hyphens}{url}  

\usepackage[hyphens]{url}
\usepackage{booktabs}  
\usepackage{multirow}
\usepackage{xspace}
\usepackage{changepage} 
\usepackage{flushend} 
\usepackage{pifont}
\usepackage{graphicx}
\usepackage{subfigure}

\usepackage{tikz}
\usepackage[inline]{enumitem}
\usepackage{enumitem}

\newcommand{\squishlist}{
 \begin{list}{$\bullet$}
  { \setlength{\itemsep}{0pt}
     \setlength{\parsep}{3pt}
     \setlength{\topsep}{3pt}
     \setlength{\partopsep}{0pt}
     \setlength{\leftmargin}{1.5em}
     \setlength{\labelwidth}{1em}
     \setlength{\labelsep}{0.5em} } }

\newcommand{\squishend}{
  \end{list}  }

  {\end{list}}

\newcounter{ecount}
  {\end{list}}

\acmConference{Arxiv}{2025}{Manuscript}

\newtoggle{techreport}
\togglefalse{techreport}

\begin{document}

\title{Public DNS Resolvers Meet Content Delivery Networks: A Performance Assessment of the Interplay}
\thanks{This work was supported by \grantsponsor{}{NSF}{} through grant \grantnum{}{CNS-2219736}.}

\author{Nicholas Kernan, Joey Li, Rami Al-Dalky, and Michael Rabinovich} 
\affiliation{
	\institution{Case Western Reserve University, USA}
}
\email{{nlk39,xxl1021,rxa271,mxr136}@case.edu}

\begin{abstract}

This paper investigates two key performance aspects of the interplay between public DNS resolution services and content delivery networks -- the latency of DNS queries for resolving CDN-accelerated hostnames and the latency between the end-user and the CDN's edge server obtained by the user through a given resolution service.  
  While these important issues have been considered in the past, significant developments, such as the IPv6 finally getting traction, the adoption of the ECS extension to DNS by major DNS resolution services, and the embracing of anycast by some CDNs warrant a reassessment under these new realities.
Among the resolution services we consider, We find Google DNS and OpenDNS to lag  behind the Cloudflare resolver and, for some CDNs, Quad9 in terms of DNS latency, and trace the cause to drastically lower cache hit rates.  
 A the same time, we find that Google and OpenDNS have largely closed
 the gap with ISP resolvers in the quality of CDNs'
 client-to-edge-server mappings as measured by latency, while
 the Cloudflare resolver still shows some penalty with Akamai, and Quad9 exhibits
 a noticeable penalty with three of the four CDNs in the study, keeping
 up only for Cloudflare CDN that does not use DNS to map clients to servers. 
 Finally, in several locations, we observe IPv6 penalty in the latency of client-to-CDN-edge-server mappings produced by the resolvers.   Moreover, this penalty does not rise above typical thresholds employed by the Happy Eyeballs algorithm for falling back to IPv4 communication. 
 Thus, dual-stacked clients in these locations may experience suboptimal performance.  
\end{abstract}

\maketitle

\section{Introduction}
\label{sec:intro}

A majority of today's Internet traffic flows to users via content delivery networks (CDNs)~\cite{Cisco_VNI}, which makes the efficient operation of these platforms a critical factor in the overall quality of experience for users. Users are rerouted from origin content servers to CDNs through DNS resolution, and a number of public DNS resolution services are available to the users as an alternative to traditional ISP-supplied DNS resolvers.   
This paper examines performance implications of the interplay between content delivery networks and public DNS resolvers, focusing on two aspects: the latency of DNS queries for resolving CDN-accelerated hostnames and the latency between the end-user and the CDN's edge server obtained by the user through a given resolution service (these user-to-edge-server mappings are referred henceforth as "client mappings").  

   Several studies considered these important issues (e.g., \cite{ager2010comparing,huang2011public,otto2012content,hours2016study}) and in particular documented lower user mapping quality produced by  public resolvers.  However, the emergence of the EDNS-Client-Subnet (ECS) DNS extension \cite{rfc7871}  to help public DNS resolvers provide high-quality CDN mappings for their users, IPv6 finally getting traction \cite{CAZ+14}, and anycast being adopted by several CDNs to route client requests to the nearest edge server, warrant a reassessment of these issues under these new realities.

This paper compares the above performance characteristics offered by four popular public DNS
resolvers (Google, OpenDNS, Cloudflare, and Quad9) when they facilitate access to content delivered by four influential CDNs (Akamai, Cloudflare CDN, Edgecast, and Fastly).   All public resolvers and CDNs we
consider are dual-stack, that is, capable of communicating with their
clients over both IPv4 and IPv6.  Thus, by recruiting vantage points
that are also dual stack, we are able to directly assess any impact
the IP version may have on these performance aspects.   

Our main contributions include the following findings.  

\squishlist
\item
We find that DNS resolution services differ in their query response time when resolving CDN-accelerated hostnames. In
particular, the resolution latency of Google DNS and OpenDNS in our measurements far exceeds that
of the Cloudflare resolution service and, for some CDNs (notably Akamai), Quad9. We find that the higher latencies of these resolvers are mainly due to their lower cache hit rates (despite cache pre-warming), which partially comports with a prior study showing less cache sharing within these resolvers'
platforms \cite{randall2020trufflehunter} (and we point out where our findings diverge). Prior studies that considered performance of public
resolvers  \cite{ager2010comparing,huang2011public,otto2012content,hours2016study}
focused on comparing them to ISP resolvers rather than to each other.
A NANOG presentation
\cite{DNS_resolvers_NANOG18} and a trade article \cite{DZone} also observed some of the latency differences among public resolvers but did not uncover the causes behind them nor assessed the resolvers' performance in the context of their interplay with CDNs.  
\item
  At the same time, Google DNS and OpenDNS have
  generally closed the gap with ISP-provided resolvers in the quality of CDNs'
  client-to-edge-server mappings (as measured by latency) documented
  previously in several studies
  \cite{huang2011public,huang2011public,hours2016study}.   On the
  other hand, the Cloudflare resolver and Quad9 still exhibit a penalty in mapping latency with some CDNs in some regions, notably with Akamai.
  We trace the causes of these performance differences to whether the particular
  resolver and the CDN agree to use the ECS extension \cite{rfc7871} in their DNS
  interactions and whether (and how) a CDN employs anycast in routing
  client requests to its edge servers. 
\item
We assess the impact of the IP version choice on the latencies of both
the DNS queries and
client-to-egde-server mappings).  While multiple prior studies compared general performance of IPv4 and IPv6 (e.g., \cite{law2008empirical,nikkhah2011assessing,zhou2008ipv6,wang2005understanding}), 
they focus on communication performance, whereas both types of latency
we consider may also be affected by server platforms, potentially engineered and
provisioned differently for IPv4 and IPv6.  We find that the DNS
latency is generally unaffected by the client's choice of IP
version to interact with the DNS platform.  At the same time, client mapping
latencies present a more nuanced picture. While for most
resolver-CDN pairs and regions the mapping latencies are little
affected by the protocol choice, we do observe substantial IPv6 penalties in some
cases, most notably in Asia.  Moreover, these penalties still
do not rise above typical thresholds employed by the Happy Eyeballs
algorithm  \cite{rfc8305} for deciding to fall back to IPv4 communication. Thus, clients in
these cases who choose IPv6, or use the
Happy Eyeballs algorithm to dynamically select between
IPv4 and IPv6, for TCP communication, may experience suboptimal
performance. 

\item
While our study is based on the snapshot dataset collected in 2023, we
have deployed two applications to allow ongoing (re)assessment of key
performance characteristics we consider. One is a web portal \cite{nick-portal} that generates custom
reports on resolver performance using similar methodology to that used
in our study but based on the data collected monthly, and the other is an
application \cite{Joey-app}  that a user can run locally to compare performance of the
resolvers from the user's specific vantage point.   
\squishend
Our measurement datasets are available at \cite{our_data}.

\section{Related Work}
\label{sec:related}

Several studies investigated the impact of using DNS public resolvers on end-users' performance \cite{ager2010comparing,huang2011public,otto2012content,hours2016study}. 
While differing in methodologies, they found that ISP resolvers 
were geographically closer to end-users \cite{huang2011public}, 
and  redirected users to more proximal CDN edge servers \cite{ager2010comparing,huang2011public,hours2016study} than the public DNS resolvers considered (collectively, Google, OpenDNS, and Level 3). However, these studies were conducted  before ECS was either proposed \cite{ager2010comparing,huang2011public} or adopted by the CDN under study \cite{hours2016study}. None of them consider the impact of IPv6 on DNS behavior.  Further, they focus on comparing public resolvers performance to that of ISP resolvers rather than to each other. An early study by Otto et al. \cite{otto2012content} considered an impact of using different ECS prefixes on the quality of user redirections by Google CDN through Google DNS -- among the only platforms that adopted ECS at the time.  We compare user redirections of multiple CDNS through multiple resolution platforms under their natural ECS behavior.

A NANOG presentation
\cite{DNS_resolvers_NANOG18} and a trade article \cite{DZone} compare the responsiveness of several public resolvers but limit their analyses to only client-resolver interaction and to only the resolution of non-CDN-accelerated hostnames.  In particular, \cite{DNS_resolvers_NANOG18,DZone} do not consider a key aspect of our study -- the comparison of latencies of CDN client mappings produced by different resolution services.

While our study focuses on the ability of the resolution services to support DNS-based client mapping by CDNs, the actual client-mapping decisions are done by CDNs' authoritative DNS platforms. The structure and operation of these platforms represent inner working of CDNs, usually hidden from an external observer.  However, Schomp et al. \cite{Schomp2020AkamaiDNS} provide a rare overview of the authoritative DNS infrastructure used by Akamai.  

Chen et. al \cite{chen2015end} studied the impact of enabling ECS at Akamai on the quality of client-to-edge-server mapping, especially for the clients using public resolvers. Their results show that enabling ECS has decreased the RTT between these clients and their edge servers by 50\%, and significantly improved other metrics, at the cost of increasing the number of DNS queries from public resolvers to Akamai's authoritative DNS servers by a factor of 8. S\'anchez et. al \cite{sanchez2013dasu} found similarly significant impact of ECS on the quality of client mapping in the EdgeCast CDN for clients using Google Public DNS. Using active measurements from a specially instrumented client application, they observed the reduction in the time to obtain the first byte of content of 20-60\% for clients in North America and Western Europe and 70-90\% for clients in Oceania. \myred{De Vries et al. \cite{de2019passive} analyze ECS-enabled queries from Google Public DNS to a busy authoritative server and uncover a privacy leak where the combination of mail servers using Google Public DNS and Google using ECS with authoritative servers leads to revealing to the authoritative servers the identities of the mail servers exchanging email with each other.}  Our study shows that, with sufficient resolver footprint or an anycast-based CDN, a public DNS resolver can provide competitive client-to-edge-server latencies without resorting to ECS as we found the Cloudflare resolver and, to a less extent, Quad9, achieve this for most regions and CDNs we consider.  

Turning to the impact of IPv6 transition, Alzoubi et.al
\cite{alzoubi2013performance} studied performance implications of
unilateral enabling of IPv6 by Websites. They found no evidence of
performance penalty for doing so, although their measurements employed
coarse time granularity of 1 second. This finding was largely
confirmed by Bajpai et. al  \cite{bajpai2016measuring}, who probed
Alexa top-10K websites from 80 vantage points and 
found that, although most tested websites had higher latency over IPv6,
91\% of these sites had IPv6 latencies within 1 msec of their IPv4
counterparts. Our investigation complements these studies by
considering IPv6 impact on the client-to-edge-server latencies of 
client mappings, and finds the impact to be much more significant in certain
cases (i.e., some resolver-CDN pairs in some regions).

An initial -- unpublished -- iteration of this study was documented in
\cite{full_paper}.  The current study uses more vantage points, employs a
more systematic approach to selecting websites for measurement, and
contributes two applications for ongoing reassessment of our findings
using current data and user-specific perspective.
\vspace{-2mm}

\section{Methodology}
\label{sec:methodology}
This study considers four popular public DNS resolvers -- Google DNS,
OpenDNS, Cloudflare, and Quad9 -- and compares them with the resolvers
provided by the ISP for a given vantage point.  Tables~\ref{tbl:resolvers-IP} lists the IP addresses of public resolver in this study. 

\begin{table}[H]
\centering
\begin{tabular}{|c|c|c|} 
\hline
\textbf{Public Resolver} & \textbf{IPv4 Address} & \textbf{IPv6 Address} \\ \hline
Google & 8.8.8.8 & 2001:4860:4860::8888 \\ \hline
Cloudflare & 1.1.1.1 & 2606:4700:4700::1111 \\ \hline
OpenDNS & 208.67.222.222 & 2620:119:35::35 \\ \hline
Quad9 & 9.9.9.9 & 2620:fe::fe \\ \hline
\end{tabular}
\caption{\label{tbl:resolvers-IP} IP addresses of public resolvers in our study}
\end{table}

To assess their
performance with regard to DNS latency and quality of client mappings
by CDNs, we must select (a) CDNs to study, (b) a sample of URLs accelerated by each of
the selected CDNs, and (c) the vantage points from which to measure
resolvers' performance.

\subsection{CDN Selection}
We select several influential
CDNs in investigating the resolvers' 
effect on CDN edge server selection.  To this end, we start with
Akamai, Verizon's Edgecast, Limelight, and CenturyLink (now rebranded
as Lumen) that were frontrunners in a market research report that ranked CDNs according to their
capabilities and market strategies (\cite{IDC_MarketScape}, Figure 1)
and augment this list with Cloudflare and
Fastly as they are frequently considered in research and were ranked close behind
the leaders.  However, we were unable to find domains accelerated
by Lumen and Limelight that resolved through AAAA DNS queries (i.e.,
could be accessed over IPv6) as
necessary for our study.  Thus, this study includes Akamai, EdgeCast, Cloudflare, and
Fastly.   Since our study includes Cloudflare's both resolver and CDN
services, we refer to the former as Cloudflare-R and the latter as
Cloudflare-CDN to avoid confusion.

\subsection{Domain Selection}
To assess the quality of client-to-edge-server mappings of the target
CDNs, we select  50 Akamai-accelerated websites and five websites
accelerated by each of the other CDN on our list\footnote{We treat Akamai as a special case due to their
exceptionally large number of points of presence and because prior
research indicated that a given domain maps to only a fraction of
Akamai's edge servers \cite{sipat_thesis}.}, for the total of 65 websites
measured from each vantage point.  We justify these website sample
sizes in Section~\ref{sec:probe_selection}.

Our website selection 
procedure crawls Majestic Million sites in the declining popularity
order and chooses the first ones that satisfy the criteria below, until
the desired number of sites for each CDN has been reached.  The
criteria are that the site includes a URL that (a) is accelerated by the CDN
in question and (b) can be delivered over both Ipv4 and IPv6
protocols.   Specifically, we follow
the resolution chain of CNAMEs for each Majestic domain and the domains  
embedded or linked within the webpage accessible at the Majestic domain, looking for suffixes known to belong to a CDN in our study
(for example, domains ending in '.edgekey.net' for Akamai).
Once found, the ``terminal'' CNAME is resolved using both A and AAAA
queries through each public resolver under study to verify the IPv4
and IPv6 support and reachability.  Finally, after selecting our
websites as above, we choose one such CNAME
per site (the first valid one we encountered) for subsequent measurements.  In the rest of the paper, we call this CNAME, interchangeably, the
website or domain accelerated by the corresponding CDN, or simply the
corresponding CDN's website (e.g., Akamai website) for brevity.

\subsection{Vantage Point Selection}
\label{sec:probe_selection}

We employ RIPE Atlas measurement platform \cite{RIPE_Atlas} for this
study, using all of its suitable probes as vantage points.  The
suitable probes include those that (a) support measurements over
both IPv4 and v6, (b) are either anchors or hardware probes with versions v3 or later (since
older versions of hardware probes were found to unpredictably inflate measured
latencies \cite{holterbach2015,bajpai2015}), and (c) for probes in over-represented regions (the US and
Europe excluding Russia), are listed as maximally stable (i.e.,
marked with 90-day stability tags)
for both IPv4 and v6\footnote{In other regions, we try to use
 probes regardless of their stability in an attempt to increase the number of
 vantage points.}. In total, 2056 probes, representing 97 countries and
959 autonomous systems, satisfied the above conditions. However, not
all these probes could be used for our study as described below.

\noindent {\bf Probes Using ISP Resolvers}

While probes can always be asked to
conduct a DNS resolution through an arbitrary public resolver, we
cannot request a probe to use the ISP-provided resolver unless the
probe already uses such a resolver by default, since we do not know
what this resolver would be.
Thus, to compare the performance of public and ISP-provided resolvers, we need
to identify probes that use the latter by default.   Note that, in
deviation from our preliminary study \cite{full_paper}, we would like to
detect ISP-provided resolvers in probes' local resolver
lists, regardless of whether or not the resolver that ultimately
interacts with authoritative name servers (AKA the ``egress resolver'')
is public.  Indeed, whether
or not the ISP-provided resolver involves public resolvers in their
resolution process is in internal ISP decision, and we would like to
include such resolvers in comparing their performance with scenarios when a client
machine is explicitly configured to use a public resolver.  

Unfortunately, RIPE Atlas does not provide information on probe's DNS
configuration.  However, it
does offer a DNS measurement using every local resolver
on the probe's list.  We therefore 
detect probes that use ISP-provided resolvers by requesting each probe
to conduct such DNS measurements to special domain names provided by
Akamai, ``whoami.ipv4.akahelp.net'' and ``whoami.ipv6.akahelp.net''
\cite{special-hostname}.  The results include responses obtained through every
local resolver, along with the IP address of the local resolver 
providing the response.  For 
resolvers with a public IP addresses, we do not make use of the data
from Akamai responses but simply infer
that a resolver is ISP-provided if it belongs to the same autonomous system as the probe
as determined by Team Cymru \cite{cymru} from both the probe's and resolver's IP addresses.  For resolvers with private
IP addresses, we must exclude situations where a probe uses the wifi
router as a local resolver (a common 
scenario) while the wifi router is configured to use a public -- rather
than ISP-provided -- resolver.  Consequently, for private addresses
only, we do use the IP address of the egress resolvers provided in
Akamai's responses to queries for the special names above, and assume
a match if the egress address belongs to probe's autonomous
system.  This approach conservatively excludes an IPS-provided
resolver with a private IP address if it
then employs an external resolver for upstream DNS interactions.   This materially affect our findings as only 6 probes in total were excluded for this reason, out of 500 probes with listed IP addresses and private resolvers.  

To be usable in comparing the performance of ISP-provided and public
resolvers, a probe's list of local resolvers needs to include at least
one IPv4 and at least one
IPv6 address that are determined to be ISP-provided using the procedure
above. We find 494 such usable probes; they are
referred to as ``ISP probes'' below.  

\subsection{Measurement Collection and Processing}
\label{Sec-Measurements}
We assess the DNS and client-to-edge-server latencies by conducting the
following measurements from
each probe for each selected website. Our initial measurement campaign was conducted on January 26-30, 2022
and the fill-in measurements describe below took place on February 13-14, 2022. 

\noindent {\bf DNS latency}. We measure DNS query latencies of
the four public resolvers under study by querying for an
IPv4 address (A-type query) and an IPv6 address (AAAA-type query)
through, respectively, the IPv4 and IPv6
address of each resolver. For the ISP
probes, we additionally conduct DNS measurements through 
ISP-provided resolvers.   ISP-provided resolvers include private
addresses, which RIPE Atlas does not allow to be directly specified as
a resolver to use. Thus, we leverage RIPE Atlas's ability to perform DNS measurements through probes'
entire resolver lists, thereby receiving a list of responses for each
probe (as a single RIPE Atlas result).  We only use the response of the first IPv4 (resp., IPv6) resolver on the list that was 
determined to be ISP-provided as described above.  

Since we are interested in the responsiveness of the
resolution services themselves, we attempt to factor out the the
uncertainties due to
the unpredictable state of caching at the time of measurement by
putting all resolution services on the level playing field with regard
to caching.  Thus, to obtain a DNS latency reading, we schedule an initial
measurement to prewarm the cache, followed by three 
back-to-back DNS measurements.  We wait 15 seconds between scheduling the prewarming and
follow-up measurements to ensure that Ripe-Atlas will schedule the
prewarming one first.  Since in practice these measurements still
occasionally occur out of order, we use the median latency of the three measurements with the
latest timestamps as the latency reading.   

While cache rewarming provides the resolution services with an equal opportunity to fill their
cache, as we will see (Section~\ref{sec:DNS_Latency_Diffs}),
some resolution services make better use of this opportunity than the
others.  However, it is fair to factor these differences into the
overall performance assessment of their platforms.  We should
emphasize that our methodology characterises DNS latency of a query
with an opportunity for a cache hit and, 
separately, of cache hits and misses.  As an active measurement study,
our investigation is unable to shed light on blended DNS latency
observed by a user in the course of their typical Internet
activities. 

\noindent {\bf CDN mapping latency.}
We assess the quality of the obtained CDN mappings for a given probe by
measuring the latency of the TCP handshake between this probe and the CDN edge
server, using the IPv4 and IPv6 addresses obtained in the
corresponding DNS measurement. Specifically, we take the first IPv4
(resp., IPv6) address from the first
non-prewarming DNS response to the A (resp., AAAA) query as the edge
server selected for this probe for this website. We then perform three
downloads of the SSL certificate from the assigned CDN edge server --
the only standard measurement in RIPE Atlas over TCP -- 
and take the median handshake RTT (time between sending the SYN and
receiving the SYN/ACK segments) for the result.  We refer to this metric as {\em mapping
  latency} below. We interchangeably call assignments of clients to
IPv4 (resp., IPv6) edge servers as IPv4 or A (resp., IPv6 or AAAA)
mappings.

\noindent {\bf Data processing.}
Our measurements are conducted from each probe employed in the
experiment, for each resolver under study, 
using each website selected to represent each CDN.  After the
measurement collection, we consider a given  (probe,
website, resolver) combination to
  have a {\em usable measurement set} if it includes at least three (out of four)
  DNS measurements and all 
  three SSL measurements.  Indeed, a set with only three successful
  DNS measurements will include at least two measurements with a
  prewarmed cache, and since we use the median for the data point, the
  potentially slow prewarming measurement will not be selected. Regardless, the combinations with only three DNS
  measurements were so rare (1,214 combinations vs. 1,062,695
  combinations with all four measurements, or 0.114\%) that whether
  or not they are included is immaterial to our findings.

When a given (probe, website) pair has usable measurement sets for all
resolvers,  we say the probe has {\em complete results} for this website.  
For a fair comparison, we use measurements for a given (probe,
website) pair only if this probe has complete results for this
website. This ensures that the performance statistics for every
resolver reflect measurements from the same set of probes using the
same set of websites from each probe.

We then compare the performance of the resolvers by considering the
DNS and mapping latency distribution for every resolver $R$, where each
probe $P$ contributes a single latency data point obtained as follows.
First, for each website $W_{C}$ accelerated by CDN $C$ with the
complete results on probe $P$, we take the median DNS latency $L^{DNS}_{P,W_C,R}$ of
the last three individual DNS latency measurements from probe $P$
obtained by resolving domain $W_C$ through resolver $R$.  Second,
we take the median $L^{DNS}_{P,C,R}$ of the median latencies
$L^{DNS}_{P,W_C,R}$ for each website $W_C$ and use this overall median
as the DNS latency data point
provided by probe $P$ for resolver $R$ when $R$ is used to access
content accelerated by CDN
$C$.
  
\subsection{Missing Measurements}

Not all probes perform measurements successfully, for a variety
of reasons, such as probe failures or timeouts, an incomplete
measurement scheduling by RIPE Atlas (perhaps due to overloaded
probes) or because RIPE Atlas never reported the measurement result.
Out of the total 1,133,340 DNS and SSL measurements from
$(\textrm{probes} \times \textrm{websites} \times \textrm{resolvers})$ combinations, 93.77\% of the
DNS measurements had all four successfully obtained results, and out
of these, 90.56\% had all three 
successful SSL measurements.  

In an attempt to obtain the missing data, we conduct fill-in
measurements to retry every (probe, website, resolver) combination
that did not produce a usable measurement set, repeating
the full set of measurements (a prewarming DNS measurement, three
follow-up DNS measurements, and three SSL measurements). If this successfully
produces a usable set, we replace the entire initial set of results for
that combination with the new results.  
After the fill-in measurements, we had 1,046,177 usable measurement sets, for a 92.3\% success rate. 

Unfortunately, only 34 probes, with
6 ISP probes among them, produced complete results for {\em all} 65
websites. At the same time, we find
that most probes do have complete results for {\em most} websites, with 1717
probes, including 428 ISP probes, having complete results for at least 30 out of 50
websites for Akamai and at least 3 out of 5 websites for every other CDN.  
Thus, we use the results from these probes for our analysis.  To
consider the effect of using fewer websites per CDN, we take the few
probes that we have with complete
results for a given CDN and examine the
difference in latency distributions for this CDN obtained from these
probes when using all 5
(50 for Akamai) websites vs. only 3 (30 for Akamai) websites
selected either at random for each probe or to have the most probes
with complete results.

\begin{figure}
    \centering
    \subfigure[DNS latencies; p=0.99999, 0.99999]{\includegraphics[width=0.7\columnwidth]{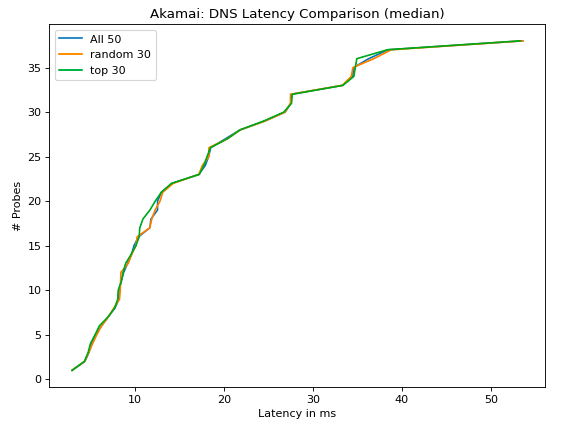}}
    \subfigure[Mapping latencies; p=0.99999, 0.98664]{\includegraphics[width=0.7\columnwidth]{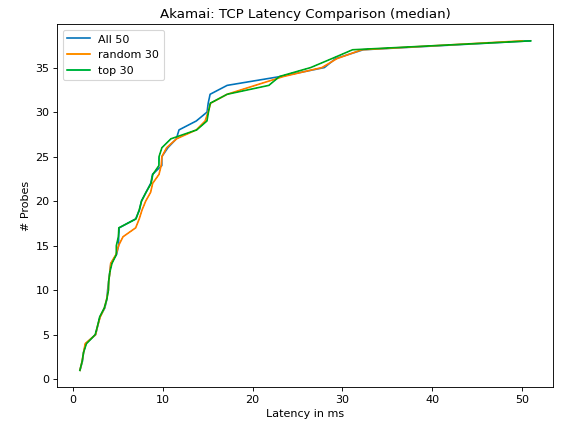}}
    \caption{Akamai latency distributions using all 50 websites
      and 30-website subsets, with p-values of K-S tests of distribution similarity.}
\label{subset-comparison}
\end{figure}

Figure~\ref{subset-comparison} shows the CDFs of DNS and mapping
latencies obtained from the 38 probes that have complete results (including both IPv4 and v6 data points) for
all 50 Akamai websites.  The three curves in each case almost entirely
overlap, providing a visual indication that choosing either a random or a most commonly
successful subsets of 30 websites does not distort the results.  This
is further corroborated by the two-sample Kolgorov-Smirnov tests, where
extremely high p-values computed for the full 50-website set and each
of the 30-website subsets are consistent with the null hypothesis that
the values used in each plot come from the same distribution.

We repeated this analysis for every non-Akamai CDN using
subsets of three randomly selected websites. 
  In each case, the plots virtually overlap as in Figure
  \ref{subset-comparison}\footnote{
We also checked the distributions of the mean latencies and found them somewhat more divergent, and also jittery, for smaller subsets of Akamai websites.  We traced this to two laggard websites in our sample,  www1.wdr.de.v1.edgekey.net. and www.dhl.de.edgekey.net, with consistently high latencies exceeding 100ms. Using medians removes the effect of such outliers}.

  This shows that using a sample of three websites per non-Akamai CDN
  and 30 websites for Akamai is sufficient for producing
  representative results that
  are unlikely to be biased -- since increasing the numbers of
  websites by 67\% does not affect the results.  Consequently, we use
  the 1717 probes (including 428 ISP probes) that produced complete
  results for at least this many websites.  
Note that these probes
provide complete results for {\em at least}, but sometimes more than,
3 (30 for Akamai) websites. However, as mentioned earlier, each probe still contributes only 
one value per resolver -- the average over all websites with complete
results on this probe -- to our 
latency distribution analysis.  

\begin{figure}[tb]
    \centering
          \includegraphics[width=0.9\columnwidth]{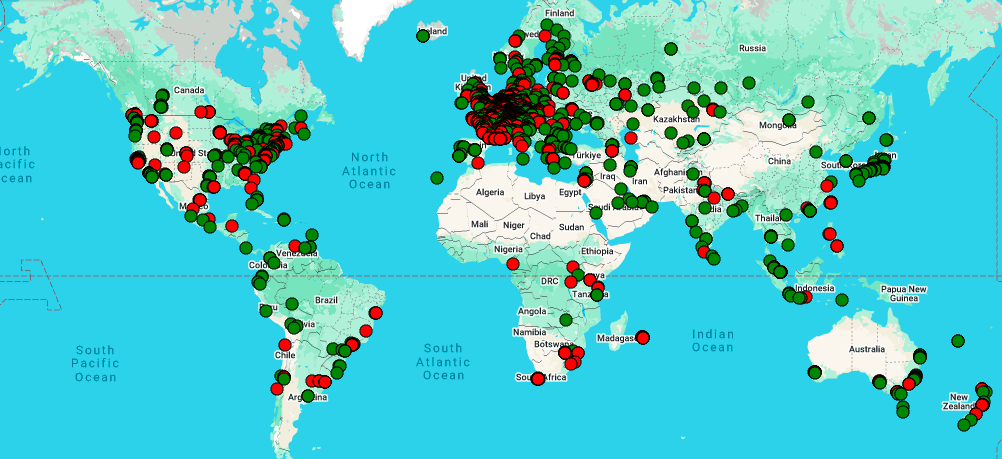}
          \vspace{-1em}
      \caption{RIPE Atlas probes used in the study.  ISP Probes are
        depicted in red.} \vspace{-2mm}
      \label{fig:probes_dist}
    \end{figure}
    
 The 1717 probes  included 1092
  in Europe, 271 in N. America, 212 in Asia, 45 in S. America, 65 in
  Oceania, and 32 in Africa.  Among these probes, there are 428 ISP probes, including 312 in
  Europe, 63 in N. America, 19 in Asia, 13 in Oceania, and 11 in
  Africa as shown in Figure~\ref{fig:probes_dist}.

\section{DNS Response Time}
\label{sec:DNS-latency}

\begin{figure*}
    \centering
    \subfigure[DNS latencies for Akamai sites.]{\includegraphics[width=0.7\columnwidth]{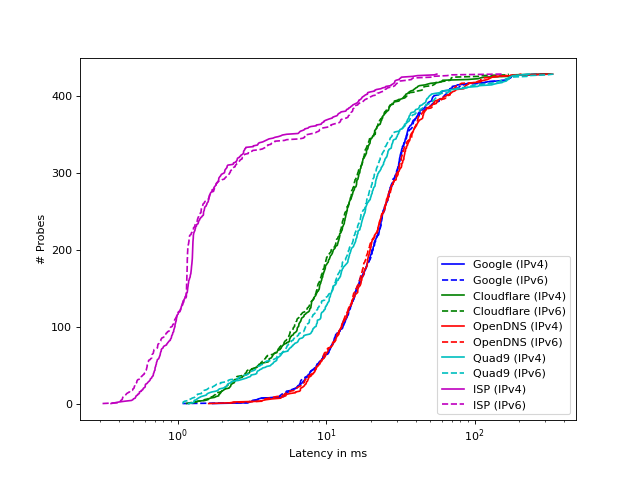}}
    \subfigure[DNS latencies for Cloudflare-CDN sites.]{\includegraphics[width=0.7\columnwidth]{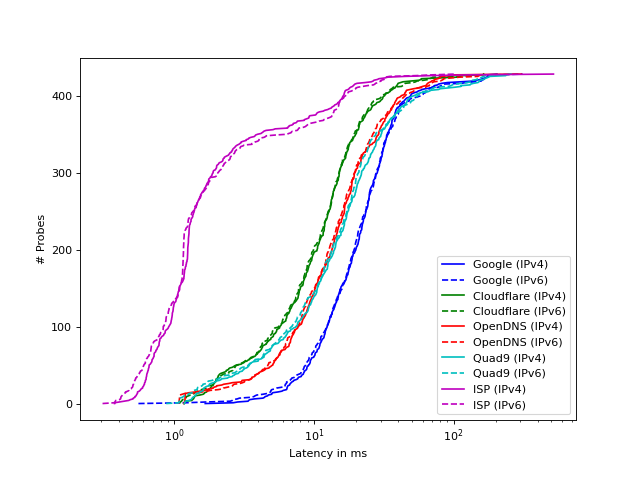}}
    \subfigure[DNS latencies for Edgecast sites.]{\includegraphics[width=0.7\columnwidth]{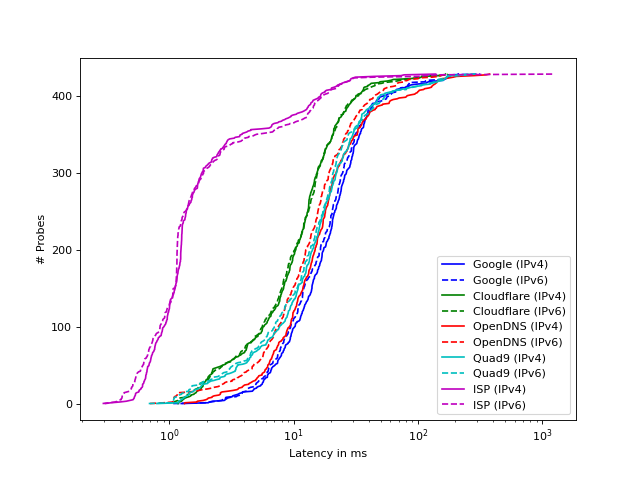}}
    \subfigure[DNS latencies for Fastly sites.]{\includegraphics[width=0.7\columnwidth]{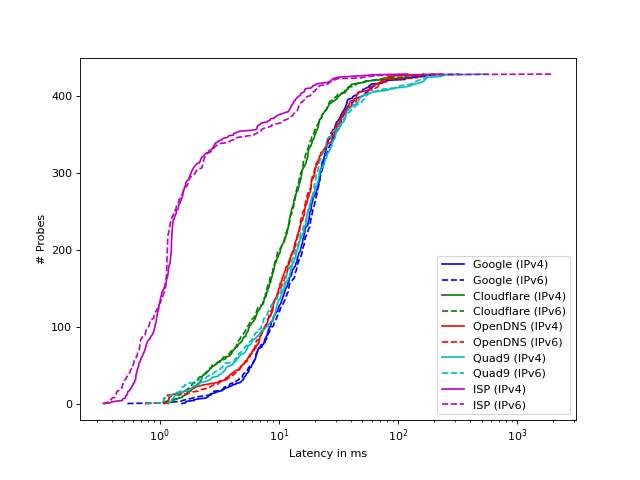}}
\caption{Distributions of DNS latencies from 428 ISP probes.}
\label{dns-latencies-median}
\end{figure*}

\begin{table*}
\small
\begin{tabular}{|c|c|c|c|c|c|c|c|c|c|c|} 
\hline
\multirow{2}{*}{CDN} & \multicolumn{2}{|c|}{ISP} & \multicolumn{2}{|c|}{Google} &  \multicolumn{2}{|c|}{Cloudflare-R} & \multicolumn{2}{|c|}{OpenDNS} & \multicolumn{2}{|c|}{Quad9}\\ \cline{2-11}
& IPv4 & IPv6 & IPv4 & IPv6 & IPv4 & IPv6 & IPv4 & IPv6 & IPv4 & IPv6\\ \hline

Akamai & 1.27 & 1.19 & 21.64 & 21.88 & 12.22 & 11.86 & 20.86 & 20.57 & 15.91 & 15.24 \\ \hline
Cloudflare-CDN & 1.26 & 1.17 & 21.07 & 20.18 & 11.13 & 10.98 & 14.06 & 13.79 & 14.83 & 14.49 \\ \hline 
Edgecast & 1.26 & 1.17 & 19.04 & 17.56 & 11.28 & 11.25 & 15.52 & 13.60 & 15.26 & 14.65 \\ \hline
Fastly & 1.26 & 1.17 & 15.94 & 16.55 & 11.26 & 11.13 & 14.20 & 14.12 & 15.19 & 14.91 \\ \hline 
\end{tabular}

\caption{Median DNS latencies from 428 ISP probes (ms).}
\label{Global-DNS-latencies}
\end{table*}

We consider the overall distribution of DNS response times
across all our vantage points, which reflects the aggregate
performance trends for various resolution services and protocols.

Figure~\ref{Global-DNS-latencies} plots the
cumulative distribution of DNS latencies for the CDNs under study.
To fairly compare public resolvers to ISP resolvers, these
distributions only use the measurements from the 428 ISP
probes.\footnote{Analogous plots for just the public resolvers using
    all probes are available in \cite{nick_thesis}; they generally show the same trends in relative performance
    among the public resolvers.}
For convenience, Table~\ref{Global-DNS-latencies} lists the median values for
  all resolvers, CDNs, and protocols. 
We make the following observations. 

\squishlist
 \item ISP resolvers respond much faster than any public resolver,
   often by an order of magnitude. As Table~\ref{Global-DNS-latencies}
   shows, median latencies of the ISP resolvers are just over 1 ms for
   every CDN, but always at least 10 ms for public resolvers. 
 \item Among public resolvers, Cloudflare-R consistently has the lowest
   latencies while Google consistently has the highest. Cloudflare-R's
   median latencies across all CDNs and IP versions are in the range
   of 10.98 - 12.22 ms, while Google's range is 15.94 - 21.88
   ms. Thus, as a global median the fastest public resolver is about
   1.5 to 2 times faster than the slowest. Quad9 and OpenDNS's latency
   is in the middle, and their performance is generally
   similar to each other with all CDNs except for
   Akamai, where Quad9 is preferable while OpenDNS exhibits similar
   latency to Google.  
 \item Every resolver has similar performance for A queries over IPv4
   and AAAA queries over IPv6
   IPv6. Thus, there is no IPv6 performance penalty in DNS
   resolutions; if anything, median latencies for IPv6 queries are
   somewhat lower (however slightly) in many resolver-CDN cases. 
\squishend

Comparing with our preliminary study 
of a few years ago \cite{full_paper}, we observe reduced DNS latencies across the
board. Some differences in methodology notwithstanding, we believe the 
reduction in DNS latencies reflects a real trend, potentially due to
a combination of factors such as expanded footprints of the resolution
platforms and reduced access network latencies over the past few years.

\subsection{On Causes for DNS Latency Differences}
\label{sec:DNS_Latency_Diffs}
While ISP resolvers can be expected to provide lower latency as they
typically situated close to their clients, we would like to understand possible reasons behind latency
differences of public resolvers. Recall that our methodology provides
all resolvers with {\em equal
opportunity} to prewarm their cache, but how they {\em use} this opportunity
depends on how they manage their distributed platforms.  Indeed,
prior work \cite{randall2020trufflehunter} has considered the same public resolvers
we use in our study and identified substantial differences in their
caching behavior.  The data below provides a strong indication that
the difference in the DNS 
cache hit rates plays a major role in DNS latency differences we
observe. Since we focus on comparing public resolvers with regard to
their caching effectiveness, we include results from all 1717 probes
with complete results in our analysis.  We also include the results
for the ISP resolvers for reference.

To assess the hit rate, we obtain the authoritative TTL of the
responses by directly querying the authoritative DNS servers of the
CDN services used by our target websites\footnote{20s for
  Akamai, 30s for Fastly\footnote{While there is a confirmation of this TTL value from Fastly's documentation (https://www.fastly.com/blog/best-practices-protecting-your-domain/), its ADNS responses at the time of this writing show that their TTL values have changed, with four of our domains returning TTL of 60s and one 300s.}, 300s for Cloudflare-CDN, and 3600s for
  Edgecast.}. Then we consider a response
to be a cache hit (resp., miss) if its TTL is equal to (resp. lower
than) the authoritative\footnote{Except for
Google, which  decrements authoritative TTL by 1 sec {\em before}
serving the authoritative response on a miss \cite{google_cache}. So
for Google we detect a cache miss if the response TTL is  one second
less than the authoritative.}. The same response that provides 
the latency is used to make the cache hit or miss determination.

\small
\begin{table*}
\begin{tabular}{|c|c|c|c|c|c|c|c|c|c|c|c|c|c|} 
\hline
\multirow{3}{*}{CDN} &
\multirow{3}{*}{Resolver} & \multicolumn{4}{|c|}{Hits} & \multicolumn{4}{|c|}{Misses} &  \multicolumn{4}{|c|}{Unknown} \\ \cline{3-14}
& & \multicolumn{2}{|c|}{IPv4} &  \multicolumn{2}{|c|} {IPv6} &
       \multicolumn{2}{|c|}{IPv4} &  \multicolumn{2}{|c|}{IPv6} & \multicolumn{2}{|c|}{IPv4} &  \multicolumn{2}{|c|}{IPv6} \\ \cline{3-14}
& & Rate& Latency & Rate & Latency & Rate & Latency & Rate
                & Lat & Rate & Latency & Rate & Latency \\ 
                     & & (\%) & (ms) & (\%) & (ms) & (\%) & (ms) & (\%) & (ms) & (\%) &
                                                                   (ms)
  & (\%) & (ms) \\ \hline \hline
\multirow{5}{*}{Akamai} & ISP & 80.26 & 1.28 & 79.69 & 1.19 & 14.44 &
                                    18.64 & 14.9 & 18.68 & 5.3 & 2.35 & 5.41 & 2.97 \\ \cline{2-14} 
 & Google & 24.14 & 13.31 & 23.46 & 13.92 & 75.66 & 22.3 & 76.32 & 22.45
                                                                & 0.2 & 35.0 & 0.22 & 22.01 \\ \cline{2-14}
 & Cloudflare-R & 75.73 & 9.55 & 75.2 & 10.09 & 24.07 & 37.29 & 24.58 & 37.22 & 0.21 & 32.79 & 0.22 &7.17 \\ \cline{2-14} 
 & OpenDNS & 12.59 & 11.22 & 13.05 & 12.44 & 87.2 & 22.23 & 86.74 & 22.4 & 0.21 & 33.12 &0.21 &8.43 \\ \cline{2-14} 
 & Quad9 & 80.59 & 14.44 & 78.92 & 13.83 & 19.19 & 25.71 & 20.86 & 26.37 & 0.22 & 33.97 & 0.22 &21.07 \\
  \hline \hline

\multirow{5}{*}{Cloudflare-CDN} & ISP & 95.56 & 1.27 & 94.35 & 1.18 & 3.82
                               & 18.33 & 5.03 & 19.9 & 0.62 & 1.21 & 0.62 & 1.19 \\ \cline{2-14} 
 & Google & 33.14 & 13.12 & 33.8 & 13.9 & 66.86 & 21.24 & 66.2 & 21.16
                               &  0.0 & N/A & 0.0& N/A \\ \cline{2-14} 
 & Cloudflare-R & 95.71 & 9.07 & 96.14 & 9.47 & 4.29 & 6.82 & 3.86 & 6.94 & 0.0 & N/A & 0.0 &  N/A \\ \cline{2-14} 
 & OpenDNS & 89.48 & 12.0 & 90.33 & 12.05 & 10.52 & 13.97 & 9.67 & 12.39 &0.0 & N/A &  0.0  &  N/A \\ \cline{2-14} 
 & Quad9 & 96.57 & 13.48 & 96.46 & 13.33 & 3.43 & 24.86 & 3.54 & 24.39 & 0.0 & N/A & 0.0 &  N/A \\ 
  \hline \hline

\multirow{5}{*}{Edgecast} & ISP & 98.8 & 1.27 & 98.29 & 1.19 & 1.2 & 27.69 & 1.71 & 26.14 &  0.0 & N/A & 0.0 & N/A \\ \cline{2-14} 
 & Google & 43.17 & 13.77 & 53.27 & 13.9 & 56.83 & 19.94 & 46.73 & 20.45 & 0.0 & N/A & 0.0 &  N/A \\ \cline{2-14} 
 & Cloudflare-R & 96.25 & 9.14 & 96.66 & 9.69 & 3.75 & 19.02 & 3.34 &                 16.57 & 0.0 & N/A & 0.0 &  N/A  \\ \cline{2-14} 
 & OpenDNS & 28.65 & 12.59 & 94.59 & 12.07 & 71.35 & 15.5 & 5.41 & 14.92 & 0.0 & N/A & 0.0 &  N/A \\ \cline{2-14} 
 & Quad9 & 99.4 & 13.91 & 98.56 & 13.44 & 0.6 & 92.82 & 1.44 & 37.93 & 0.0 & N/A  & 0.0 &  N/A \\ 
\hline \hline 
 
\multirow{5}{*}{Fastly} & ISP & 92.31 & 1.26 & 90.17 & 1.17 & 3.93 & 18.66 & 5.88 & 20.96 & 3.76 & 1.26 & 3.95 & 1.62 \\ \cline{2-14} 
 & Google & 82.65 & 12.49 & 78.57 & 12.87 & 17.19 & 21.93 & 21.34 & 22.5 & 0.15 & 35.65 & 0.09 & 43.42 \\ \cline{2-14} 
 & Cloudflare-R & 85.31 & 8.71 & 84.95 & 9.16 & 14.55 & 11.7 & 14.97 & 12.06 & 0.14 & 35.26 & 0.08 & 42.49 \\ \cline{2-14} 
 & OpenDNS & 83.93 & 11.61 & 76.72 & 11.71 & 15.92 & 15.59 & 23.17 & 14.07 & 0.15 & 35.99 & 0.1 & 43.34 \\ \cline{2-14} 
 & Quad9 & 88.14 & 13.51 & 86.72 & 13.13 & 11.71 & 24.65 & 13.18 & 26.59 & 0.14 & 21.54 & 0.1 & 43.41 \\ \hline 
\end{tabular}
\caption{DNS cache hits and misses from 1717 probes (428 for ISP resolvers)}
\label{hit-rate-table2}
\end{table*}
\normalsize

Table~\ref{hit-rate-table2} lists the hit and miss rates of the different
resolvers for different CDNs, along with the median latencies of the
queries that hit and missed in the cache.  
  The ``Unknown'' columns in
both tables 
reflect rare cases where returned TTL values were above the authoritative
TTL value for that CDN, making it impossible to assess whether the response
is a hit or miss. We have no firm explanation for this behavior, although
past studies found that some resolvers override TTL
values before returning them to the client
\cite{ttl-violations,schomp2013measuring}, reporting this behavior particularly for local resolvers of
RIPE Atlas probes \cite{ttl-violations}.  Indeed, we find that these unexpected TTLs occur
far more frequently with local ISP resolvers as
Table~\ref{hit-rate-table2} shows.  It could also be that CDNs assign
higher TTLs to a small fraction of queries (e.g., for experimental
purposes), but this would not explain the difference in the occurence rate of
these TTLs we observe through ISP and public resolvers.   While we do
see different rate of inflated TTLs from different CDNs, 
which would seem to counter the conjecture that it is due to 
resolvers' TTL substitution, we see that inflated TTLs are more
frequent for the CDNs with the higher the authoritative TTLs. Note that TTL substitutions
are less likely to be detected for CDNs
with higher authoritative TTLs, simply because the substituted TTL may still
be within the expected range, which may explain the CDN differences in
this regard.  In the extreme case, 
with Edgecast's authoritative TTL of 3600, no TTL 
substitution is reported because even the highest TTL
substitutions observed across the CDNs were 3600s and still fall into the expected
range for Edgecast.  As a corollary, an undetected TTL
substitution may be characterized as either a hit or a miss
(depending on the ultimate TTL value) incorrectly.   Given the limited
occurrences of unexpected TTLs even for Akamai with its low
authoritative TTL of 20s, we believe any resulting inaccuracy is
immaterial for the discussion below.

Table~\ref{hit-rate-table2} illuminates stark differences in cache
effectiveness among the public resolvers.  Cloudflare-R and Quad9 have
consistently high hit rates for every CDN, in the 75-99\% range.
Google and OpenDNS's hit 
rates vary wildly with different CDNs, ranging from as low as around
13\% for OpenDNS with Akamai (and under 25\% for
Google with Akamai) to over 80\% with Fastly (comparable to Cloudflare-R
and Quad9).  Google in particular exhibits drastically lower hit rates
than Cloudflare-R and Quad9 with all CDNs except for Fastly, while
OpenDNS, in addition to Fastly, has competitive hit rates for
Cloudflare-CDN as well.   These findings indicate that the dynamic cache
rotation observed in our preliminary study to affect Google
and OpenDNS's hit rates still impacts
their performance.  These findings are also consistent with the
observations in both our preliminary study and the Trufflehunter study
\cite{randall2020trufflehunter}  that Cloudflare-R maintains a shared frontend cache
architecture for a given PoP. At the same time, our findings with
regard to Quad9 diverge from the Trufflehunter study, which concluded
that Quad9 caching mechanism is similar to OpenDNS (and 
essentially to Google as they were only found to differ in how they handle
TTLs of cached records but not hits and misses themselves).  In
contrast, we find that Quad9's hit rate is consistently high, on par
with Cloudflare-R's.   Moreover, hit
rate differences across CDNs indicate potentially differentiated handling of
queries for different CDN hostnames.  While further investigation is
needed to better understand these differences, they show the danger of
drawing generalized conclusions from individual domain measurements.

Notably, Cloudflare-R shows no miss penalty with Cloudflare-CDN websites,
with the latencies of misses being if anything lower than those of hits.
This is probably the result of processing misses within its own
platform.  OpenDNS also presents an interesting behavior: while IPv4 and v6 hit rates
for the same resolver are generally very close, indicating similar
caching strategies used for both versions, OpenDNS with Edgecast shows
a major disparity between the IP versions (28.65\% vs. 94.59\%).  We
have no explanation for this finding.  

These differences in resolvers' caching behavior correspond to the overall latency
differences in Figure~\ref{Global-DNS-latencies} and
Table~\ref{hit-rate-table2}, with impact from individual hit and miss
latencies playing a secondary role. In particular, Google's low hit 
rates with Akamai, Cloudflare-CDN, and Edgecast translate into lagging
performance across these CDNs, while Cloudflare-R's high hit rates with
these CDNs make it the highest-performing public
resolver.  OpenDNS's higher IPv6 hit rate with Edgecast (compared with
its IPv4 hit rate with the same CDN) leads to its lower overall latencies
over IPv6 than IPv4 (Figure~\ref{hit-rate-table2}c).  Even the very
high IPv4 miss latency of Quad9 with Edgecast is 
successfully masked by its high hit rate, produce a competitive
overall latency distribution in Figure~\ref{hit-rate-table2}c.  Only
when hit rates are comparable (e.g., in the case of Fastly or
Cloudflare-CDN other than with Google), it is the
individual latencies that separate the overall latencies of the public
resolvers.  Finally, the combination of consistently high hit rates with
extremely low hit latencies allow ISP resolvers to continue to
hold a substantial performance advantage over all public resolvers (as
reported in past studies \cite{huang2011public,otto2012content,hours2016study}).

\section{Regional DNS Latency \label{Regional DNS Latency}}

Because our global latency analysis can be skewed by disproportionally large number of probes in Europe and North America, we compare resolver latencies for separately individual continental regions.  

Table \ref{regional-dns-median-table} lists median DNS latencies observed in different regions through the resolvers under study. Note, we include all 
1717 responsive probes in our regional analysis as we do not have a substantial number of responsive ISP probes in each region.  Thus, the results in Table \ref{regional-dns-median-table} can be used to compare the performance of public resolvers to each other but not to the ISP resolvers: while we still provide the ISP latency results for completeness, they are not directly comparable to the public resolver results since these results are produced from different sets of probes.

\begin{table*}

\includegraphics[width=1\textwidth]{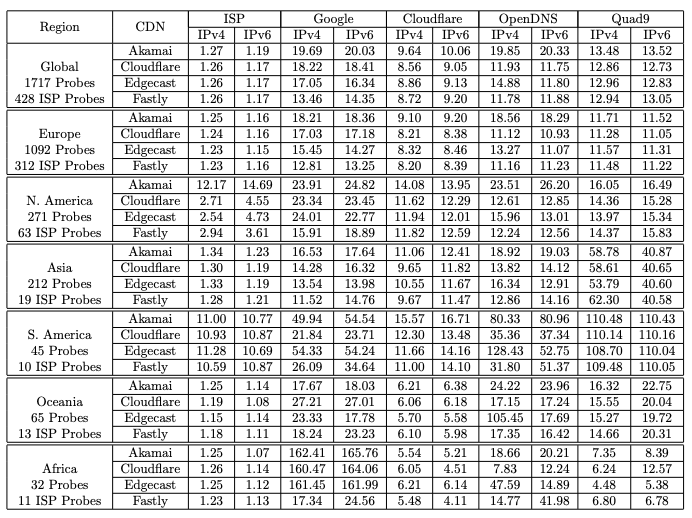}
\caption{Breakdown of median DNS latencies by continent. ISP resolver medians are provided for illustration, but we take caution in directly comparing their medians to that of public resolvers, since the public resolver medians include non-ISP probes.} 
\label{regional-dns-median-table}
\end{table*}

Many of the trends in each region resemble the global trends from Section \ref{Global-DNS-latencies}, with Cloudflare-R offering the best performance among public resolvers. However, performance differences between the resolvers in South America and especially Africa are more pronounced.

In particular, Table \ref{regional-dns-median-table} shows that in South America, the median DNS latency of Cloudflare-R was significantly lower than other public resolvers across all CDNs. Quad9 had particularly poor performance in the region, with median latencies consistently above 100 ms, up to an order of magnitude slower than Cloudflare-R, depending on the CDN and IP version. Quad9 itself realized their poor performance in the region, and announced a partnership with EdgeUno to expand their network in Latin America \cite{quad9-south-america}. Indeed, we notice in our recurring measurements that their performance has indeed improved since the announcement. 

Africa has the most dramatic performance gap noticed in our study. Table \ref{regional-dns-median-table} shows that Cloudflare-R latencies are very low, in the range of 4.11 ms - 6.14 ms across all CDNs and IP versions, while Google's median latencies are consistently above 160 except with Fastly. This can be understood by examining Google's hit and miss latencies specific to probes located in Africa, as seen in Table~\ref{tbl:google-DNS-Latencies-Africa}. 

\begin{table}[H]
\centering
\begin{tabular}{|c|c|c|c|c|} 
\hline
\multirow{2}{*}{CDN} & \multicolumn{2}{|c|}{Hit Latency (ms)} & \multicolumn{2}{|c|}{Miss Latency (ms)} \\ \cline{2-5}
& IPv4 & IPv6 & IPv4 & IPv6 \\ \hline

Akamai & 16.39 & 42.85 & 165.07 & 168.89 \\ \hline
Cloudflare-CDN & 15.27 & 19.43 & 164.99 & 168.47 \\ \hline
Edgecast & 13.88 & 25.42 & 163.38 & 165.95 \\ \hline
Fastly & 16.34 & 18.18 & 162.43 & 163.26 \\ \hline

\end{tabular}
\caption{\label{tbl:google-DNS-Latencies-Africa}Median cache hit/miss latencies for 32 African probes using Google's resolver.} 
\end{table}

As the table indicates, Google's miss latencies in Africa are above 160 ms with every CDN and IP version and up to ten times greater than their hit latencies ( particularly with IPv4). Furthermore, given Google' poor cache hit rate in Table \ref{hit-rate-table2} with every CDN except Fastly, the median DNS latencies for those CDNs are largely determined by the miss latencies. 
By comparison, cache misses do not explain Quad9's poor performance in South America: even its median hit latency was around 110 ms. 

Finally, we note that many cases of high latencies (over 100 ms) observed in our earlier study \cite{full_paper} have since been rectified. For example, across all IP versions \cite{full_paper} found Google and OpenDNS to have latencies approaching or above 100 ms in Asia, South America, Oceania and Africa (with OpenDNS in Africa exceeding even 200 ms). Table \ref{regional-dns-median-table} indicates that median latencies above 100 ms have since become exceedingly rare outside of the regions mentioned. This likely reflects an expansion of Google and OpenDNS's network, similar to Quad9's attempts to increase coverage in South America.

\section{CDN Mapping Latency}
\label{sec:mappings}

As a projected 72\% of Internet traffic is carried by CDNs
\cite{cisco-cdn-traffic}, their performance largely determines the
overall quality of the users' Internet 
experience.  Thus, an important component of evaluating DNS resolvers
is assessing the quality of client-to-edge-server mappings received
when using 
them.  We assess the mapping
quality by the network proximity of the client to the selected edge
server -- a key (although not the only) factor in CDNs' edge server selection.  As described 
in Section~\ref{Sec-Measurements}, we measure the
network proximity through the TCP handshake
latency from a vantage point to the edge server obtained in the
first non-prewarming DNS query from this vantage point to a given CDN.

\subsection{Overview}

We begin by taking a high-level view of client mappings obtained
through the public resolvers under study and comparing them to those
produced by the ISP resolvers.  Table~\ref{isp-ssl-table} gives an
overview of median mapping latencies from the 428 ISP
probes\footnote{The full distributions are shown in \cite{nick_thesis}
  and are omitted here for brevity as they only reconfirm the
  medians-based analysis.}. 

\begin{table*}
\centering
\footnotesize
\begin{tabular}{|c|c|c|c|c|c|c|c|c|c|c|} 
\hline
\multirow{2}{*}{CDN} & \multicolumn{2}{|c|}{ISP} & \multicolumn{2}{|c|}{Google} &  \multicolumn{2}{|c|}{Cloudflare-R} & \multicolumn{2}{|c|}{OpenDNS} & \multicolumn{2}{|c|}{Quad9}\\ \cline{2-11}
& IPv4 & IPv6 & IPv4 & IPv6 & IPv4 & IPv6 & IPv4 & IPv6 & IPv4 & IPv6\\ \hline

Akamai & 10.32 & 10.78 & 10.48 & 11.35 & 13.59 & 12.18 & 10.40 & 11.42 & 15.66 & 16.79 \\ \hline 
Cloudflare-CDN & 10.65 & 10.48 & 10.96 & 10.61 & 10.92 & 10.43 & 10.86 & 10.42 & 10.49 & 10.55 \\ \hline 
Edgecast & 12.71 & 12.99 & 12.80 & 12.77 & 12.71 & 13.07 & 12.82 & 12.97 & 15.37 & 14.73 \\ \hline 
Fastly & 12.78 & 13.02 & 12.67 & 13.22 & 12.82 & 13.02 & 13.08 & 13.00 & 16.59 & 17.77 \\ \hline 

\end{tabular}
\caption{Median mapping latencies with 428 ISP probes}
\label{isp-ssl-table}
\end{table*}

We can make two immediate observations.  First, there is no discernible difference between IPv4 and IPv6 mapping latencies, across all CDNs and resolvers (although see Section~\ref{sec:regional_mappings} for special cases in several regions).  

Second, Google and
OpenDNS, perform as well as ISP resolvers with all CDNs -- thus closing the gap in
CDN mapping latency reported in prior studies \cite{otto2012content,huang2011public,hours2016study}.   
Cloudflare-R is also on par with the ISP resolvers except for Akamai, where it lags behind; Quad9, however, has mapping latencies similar to the other resolvers only with Cloudflare-CDN and otherwise has higher mapping latencies. To better understand these findings we analyze each CDN separately below.

\noindent {\bf Akamai}
Akamai's mapping latencies obtained through Google and OpenDNS are on
par with those produced by ISP resolvers, whereas the latencies
provided by Cloudflare-R and Quad9
are higher.  A likely reason is
that Google and OpenDNS employ ECS in their 
interactions with Akamai while Cloudflare-R and Quad9 do not.  
Akamai uses a widely distributed platform, with over 4100 locations
in 134 countries \cite{akamai-distribution} -- an order of magnitude
more than other CDNs in this study.
The greater number of locations results in a denser distribution of
points of presence and therefore
increases the impact of ECS on mapping quality. Indeed, ECS 
allows the edge server selection to be done relative to the actual eyeballs network
location whereas without ECS the selection must use the
resolver's location as the proxy for the client.  With more points of
presense (PoPs) relatively close to each other, there is a higher chance that
the PoP closest to the resolver would be different than the
one closest to the client, leading to a suboptimal mapping.

Between Cloudflare-R and Quad9, Cloudflare-R produces lower Akamai mapping
latencies.  Since neither employs ECS, the quality of edge server
mappings in both cases depends on how well the location of the resolver
represents the eyeballs' location.  Cloudflare, which states that its network
extends to 285 cities in around 100 countries
\cite{cloudflare-locations}, has presence in more 
locations than Quad9, which claims to be deployed in around 150 locations in 90
nations \cite{quad9-locations}.  We can therefore attribute Cloudflare-R's
superior performance to Quad9 
to a larger global footprint, allowing Cloudflare resolvers to more
accurately represent eyeball locations for more clients.

\noindent{\bf Cloudflare CDN}
All resolvers produce equally good mappings for Cloudflare-accelerated
websites. This can be attributed to Cloudflare-CDN's use of global
anycast \cite{cloudflare-anycast2}. Unlike Akamai, which returns hundreds to
thousands of unique IP addresses for each website, each Cloudflare-CDN
site returns only a small number of designated IP addresses: four of
the five Cloudflare-CDN sites in our study have  only
two IPv4 and two IPv6 addresses that are returned to all probes
regardless of their location, while
the last one has only 5 of each.  Thus, Cloudflare-CDN mapping latencies
are not dependent on the resolver used, as
clients are mapped to data centers by Internet routers based on their
routing tables and not by DNS-based server selection. 

\noindent{\bf Edgecast}

\begin{figure*}
  \centering
    \subfigure[Cloudflare-CDN addresses obtained through Google]
    {\includegraphics[width=0.3\textwidth]{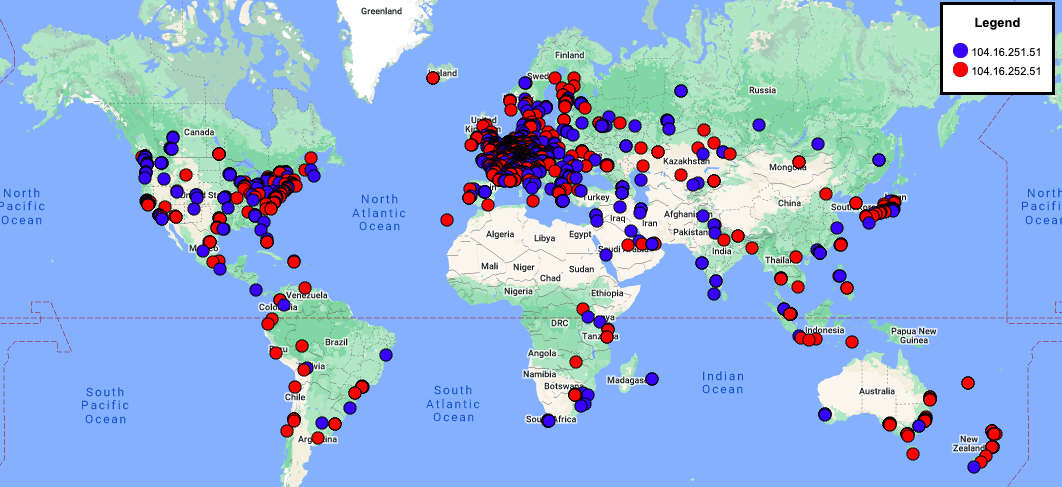}}
    \subfigure[Edgecast addresses obtained through Google]{\includegraphics[width=0.3\textwidth]{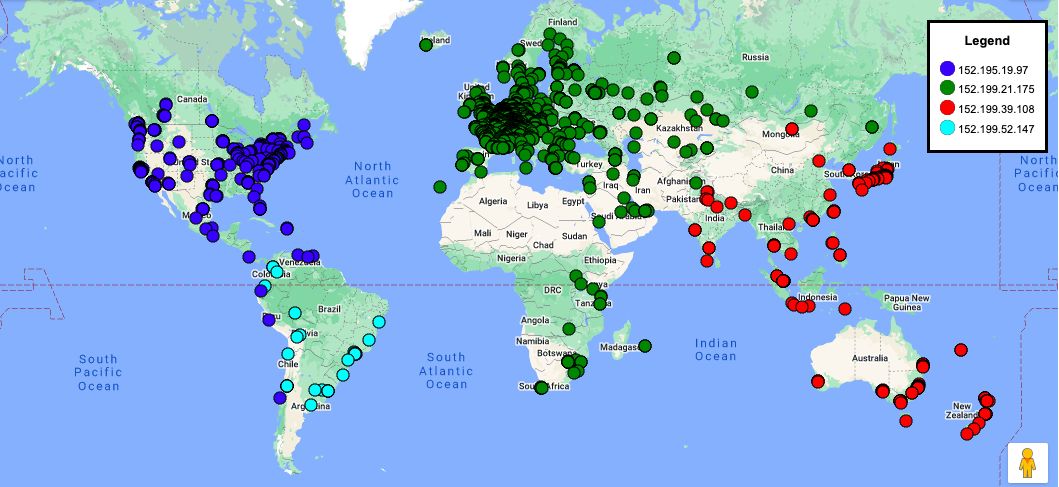}}
    \subfigure[Edgecast addresses obtained through Quad9]{\includegraphics[width=0.3\textwidth]{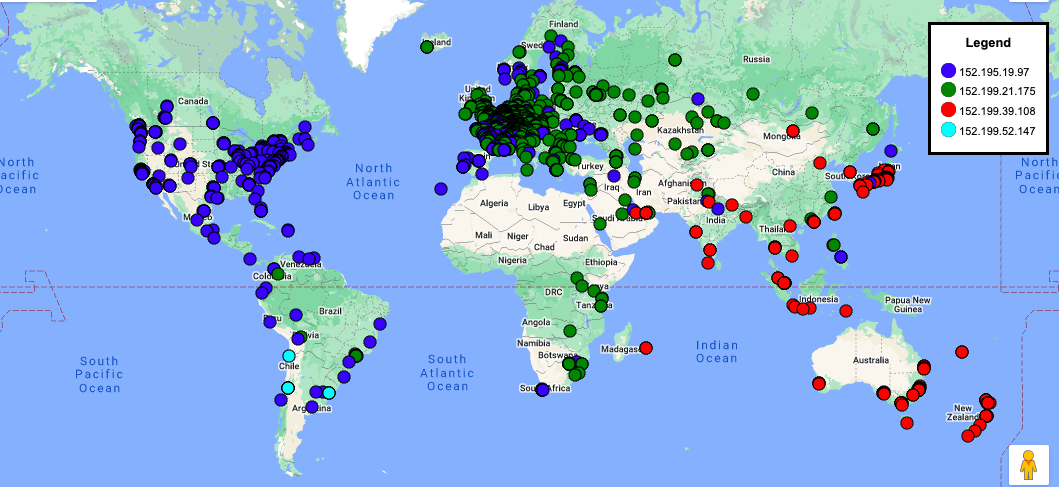}}
    \caption{Probe locations color-coded according to the IPv4 addresses obtained by
    probes for Cloudflare-CDN and Edgecast websites.}
    \label{fig:Cloudflare_Edgecast_ColorMaps}
\end{figure*}

With Edgecast-accelerated sites, Quad9 is a laggard while all other
resolvers perform on par with ISP resolvers. Similar to Cloudflare-CDN,
Edgcast uses anycast and returns only four IPv4 and IPv6 addresses
across all the probes (although unlike 
Cloudflare-CDN, this set of addresses is shared between Edgecast-accelerated
websites)\footnote{Strictly speaking, this statement is fully correct
  only with the Edgecast
  websites ending with the ``systemcdn.net'' DNS suffix.  The two
  websites with the domain suffix ``edgecastcdn.net'' in our study returned around
  50 additional IP addresses, beyond the common addresses, and one of
  the two sites only used three out of the four common addresses.
  Because 
  these extra addresses were extremely rare, with most appearing only
  once -- in contrast with tousands of occurrences of the common addresses, and
  they typically timed out when the probe attempted to connect to them
  over SSL, they did not affect the results.}. With anycast-based client mapping and only 3-4 IP addresses per
website, why would Quad9 demonstrate worse CDN mappings than the other
resolvers?  To gain an insight into this phenomenon, compare the
color-coded maps of probe locations where each probe is colored based on the IP address it
received from its DNS measurements for a given website through a given resolver. The
first non-prewarming measurement is used, since its result is the CDN
mapping we use in our measurements (see Section
\ref{sec:methodology}). In particular, Figure \ref{fig:Cloudflare_Edgecast_ColorMaps}a is
produced for Cloudflare-CDN using
site "www.propublica.org.cdn.cloudflare.net." with Google's
8.8.8.8 resolver, while Figure~\ref{fig:Cloudflare_Edgecast_ColorMaps}b and c are produced for Edgecast using site scdn1.wpc.81638.systemcdn.net with Google and Quad9.

The Cloudflare-CDN map shows that its anycast addresses are not
tied to any specific region: a mix of the same two addresses are intermixed throughout all regions. The same behavior is
observed with other Cloudflare-CDN websites and other
resolvers. Indeed, Cloudflare states that every router in every data 
center in its platform announces all its IP addresses
\cite{cloudflare-anycast2}. Thus, its mapping latencies  are not dependent on the resolver used, as clients are
directed to the topologically nearest data centers based on BGP
routes, regardless of the specific IP address obtained.

However, the Edgecast maps show that it instead uses {\em regional
anycast}.  Indeed, considering Figure~\ref{fig:Cloudflare_Edgecast_ColorMaps}b, where
probes are color-coded according to the IP addresses they obtain
through Google, there is clear regional separation, with Africa, Europe, and Western parts of Asia  directed to "152.199.21.175", the rest of Asia and Oceania to
"152.199.39.108", North America to "152.195.19.197", and South America
mostly directed to "152.199.52.147" except for some coastal locations
that are assigned to the North American address. OpenDNS
and Cloudflare-R show similar behavior (the maps are not shown as
they are essentially identical to Google). However, with Quad9 
there are far more areas of regional overlap, particularly in South
America and Europe, where North America's address of "152.195.19.197"
becomes as prevalent as the main address for that region. This likely
indicates that Quad9's lack of ECS information causes mappings to an
incorrect region, leading to the performance detriment. While Cloudflare-R 
does not employ ECS either,     
it performs on par with ISP and ECS resolvers, which we attribute to its
wider global distribution, allowing Cloudflare
resolvers to adequately represent clients' regions.

\begin{figure*}
    \centering
    \subfigure[Addresses received through Google]{\includegraphics[width=0.3\textwidth]{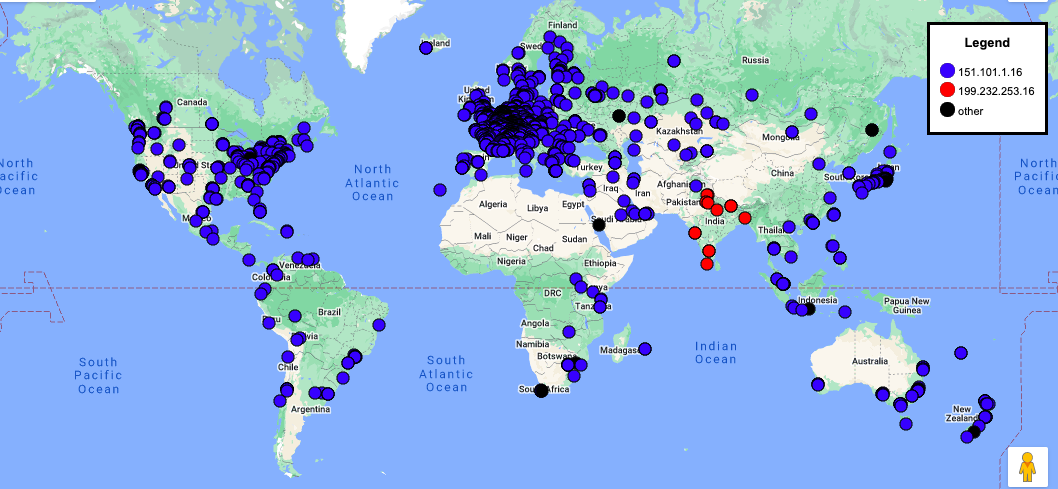}}
    \subfigure[Addresses received through OpenDNS]
    {\includegraphics[width=0.3\textwidth]{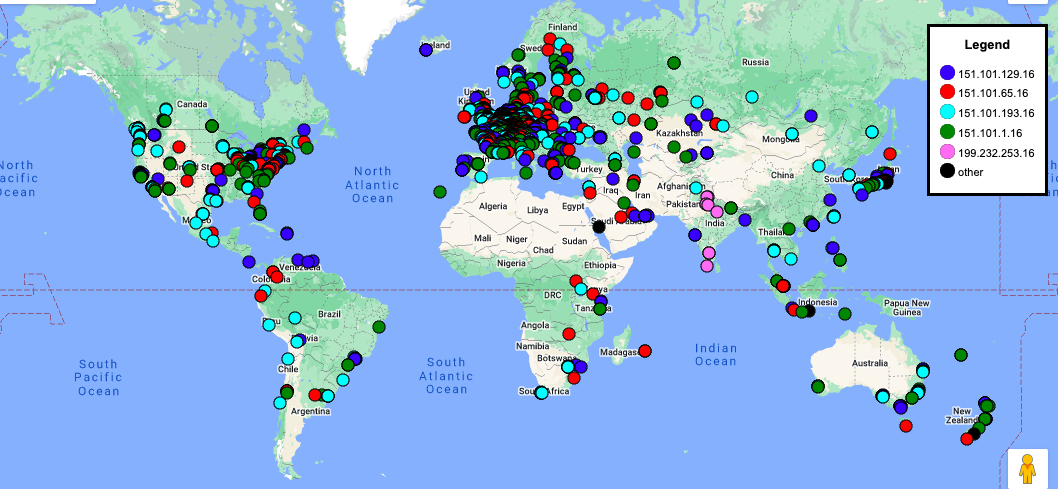}}
    \subfigure[Addresses received through Cloudflare-R]{\includegraphics[width=0.3\textwidth]{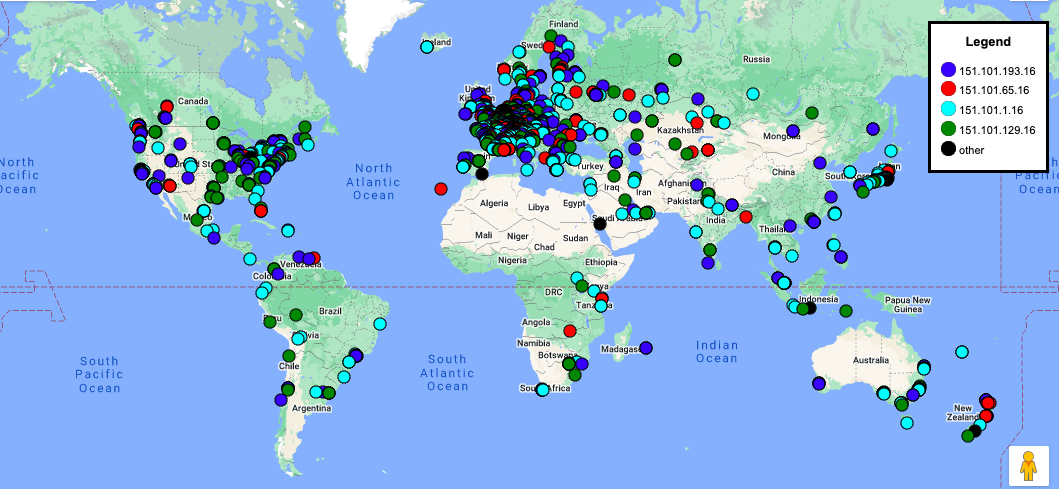}}
    \caption{Probe locations color-coded according to the IPv4
      addresses received by probes for Fastly websites}
    \label{fig:fastly-ip-distribution}
\end{figure*}

\noindent{\bf Fastly}
The Fastly mappings show a similar trend to Edgecast: Google, OpenDNS,
and Cloudflare-R perform on par with ISP resolvers, but Quad9 produces
worse mappings. We observe an interesting behavior that leads to this finding, by treating queries coming from different public resolvers differently. Fastly offers its subscribers an option to either use CNAMEs to redirect client requests to Fastly or map original subscribers' domains to Fastly's anycast addresses \cite{fastly-anycast-performance}.  However, although our methodology only employs CNAME-redirected URLs, we observe that Fastly still appears to return anycast
addresses for Google, OpenDNS, and Cloudflare-R. Indeed, IP distribution maps with
the sample Fastly site "media.amazon.map.fastly.net" reveal that a 
handful of IP addresses account for the vast majority (over 95\%) of DNS responses 
received through Google (two IPv4 and two IPV6 addresses), OpenDNS (five IPV4 and five IPV6 addresses), and
Cloudflare-R (four IPv4 and four IPv6 addresses).  Further, as
Figure~\ref{fig:fastly-ip-distribution} illustrates, these addresses
are intermixed across the regions (with the exception of one address localized to India for Google and OpenDNS). A small number of globally intermixed IP
addresses is consistent with global anycast being used for
client-to-edge-server mappings. In contrast, Quad9 exhibits a starkly
different behavior, receiving 53 IPv4 and also 53 IPv6 addresses across all the responses 
(making a color map infeasible). This indicates that Fastly does not use global anycast for DNS queries 
from Quad9, causing suboptimal mapping performance due to the lack of ECS.  

\subsection{Regional Mapping Latencies}
\label{sec:regional_mappings}

We turn to regional differences in mapping latencies produced by different resolvers.  Tables \ref{tbl:regional_akam} -- \ref{tbl:regional_fastly} show median mapping latencies for each region and each resolver when using that resolver to access a given CDN. As before, to increase the number of probes suitable for analysis, we focus on public resolvers and forgo their direct comparison with ISP resolvers.

The per-region analysis reveals several interesting points that are obscured in the global analysis by the dominance of the European and North American vantage points.  

\begin{itemize}
\item Most scenarios in Asia exhibit an IPv6 penalty in mapping latency.  Indeed, Cloudflare-CDN, Edgecast, and Fastly all show median mapping latencies to be lower for IPv4 than IPv6 with all resolvers. The IPv6 penalty in Asia is especially noticeable for Edgecast, where the penalty ranges from 2.8ms (or 37\%) for OpenDNS to 7.7ms (over 50\%) for Quad9.  Akamai  exhibits a substantial IPv6 mapping latency penalty with Cloudflare-R (6.5ms or 60\%) but not with other resolvers, where the difference between median IPv4 and IPv6 latencies are all within 10\% of each other.
\item In Africa, Fastly suffers from a substantial IPv6 mapping latency penalty across all the resolvers, with medial IPv6 latencies being 2-3 times higher than those in IPv4.  Among the rest of the CDNs, only Akamai with Cloudflare-R exhibits a significant IPv6 latency penalty of over 10ms or 80\%. Curiously, accessing Edgecast through Quad9 produces substantially -- by over 30ms or 2.7 times -- {\em lower} mapping latencies over IPv6 than IPv4. Understanding the reasons for these findings requires delving into regional differences in Internet deployments and is outside the scope of our current study. 
\item As with DNS latency, Quad9 lags far behind in the mapping latency it produced in South America, with every CDN except Cloudflare-CDN (whose mappings are intependent of the resolver employed due to its use of global anycast). Again, Quad9 has since rectified this deficiency through its partnership with EdgeUno \cite{quad9-south-america}. However, we observe Quad9 to produce disproportionally higher (as compared to global statistics) median mapping latencies in some other regions as well, specifically, in Asia with Akamai and Fastly and in Africa with Edgecast over IPv4. The differences in median mapping latency between Quad9 and other resolvers in these cases are roughly in the 30-60ms range, a rather substantial gap. 

\end{itemize}

\begin{table*}
\footnotesize
\centering
\begin{tabular}{|c|c|c|c|c|c|c|c|c|}
\hline
\multirow{2}{*}{Region} &
\multicolumn{2}{|c|}{Google} & \multicolumn{2}{|c|}{Cloudflare-R} & \multicolumn{2}{|c|}{OpenDNS} & \multicolumn{2}{|c|}{Quad9} \\ \cline{2-9}
& IPv4 & IPv6 & IPv4 & IPv6 & IPv4 & IPv6 & IPv4 & IPv6 \\ \hline

Global (1717 Probes) & 8.74 & 9.61 & 11.39 & 11.17 & 8.68 & 9.61 & 14.52 & 15.43 \\ \hline 

Africa (32 Probes) & 16.81 & 16.85 & 13.30 & 24.13 & 16.05 & 17.30 & 17.01 & 16.60 \\ \hline 

Asia (212 Probes) & 10.29 & 11.00 & 10.90 & 17.40 & 9.95 & 10.90 & 64.94 & 58.93 \\ \hline 

Europe (1092 Probes) & 8.02 & 8.75 & 11.12 & 10.28 & 8.01 & 8.65 & 12.61 & 13.29 \\ \hline 

Oceania (65 Probes) & 8.51 & 11.16 & 7.77 & 9.07 & 8.49 & 10.70 & 17.60 & 16.97 \\ \hline 

N. America (271 Probes) & 12.46 & 13.69 & 13.45 & 13.88 & 11.99 & 13.73 & 16.52 & 16.75 \\ \hline 

S. America (45 Probes) & 9.86 & 10.26 & 14.86 & 15.21 & 10.14 & 10.08 & 108.38 & 148.60 \\ \hline
\end{tabular}
\caption{Median mapping latencies for Akamai by region}
\label{tbl:regional_akam}
\end{table*}

\begin{table*}
\centering
\footnotesize
\begin{tabular}{|c|c|c|c|c|c|c|c|c|}
\hline
\multirow{2}{*}{Region} &
\multicolumn{2}{|c|}{Google} & \multicolumn{2}{|c|}{Cloudflare-R} & \multicolumn{2}{|c|}{OpenDNS} & \multicolumn{2}{|c|}{Quad9} \\ \cline{2-9}
& IPv4 & IPv6 & IPv4 & IPv6 & IPv4 & IPv6 & IPv4 & IPv6 \\ \hline

Global (1717 Probes) & 7.96 & 8.40 & 8.10 & 8.33 & 8.07 & 8.32 & 8.00 & 8.35 \\ \hline 

Africa (32 Probes) & 4.04 & 4.04 & 4.17 & 4.01 & 3.81 & 4.05 & 4.04 & 3.89 \\ \hline 

Asia (212 Probes) & 7.85 & 10.12 & 8.12 & 10.23 & 7.35 & 10.20 & 8.08 & 9.99 \\ \hline 

Europe (1092 Probes) & 7.60 & 7.86 & 7.82 & 7.87 & 7.84 & 7.89 & 7.63 & 7.86 \\ \hline 

Oceania (65 Probes) & 5.46 & 5.32 & 5.30 & 5.33 & 5.21 & 5.72 & 5.19 & 5.35 \\ \hline 

N. America (271 Probes) & 11.56 & 11.64 & 11.73 & 11.17 & 11.41 & 11.55 & 12.00 & 11.91 \\ \hline 

S. America (45 Probes) & 10.61 & 10.63 & 10.83 & 12.75 & 10.87 & 10.33 & 10.73 & 10.49 \\ \hline 
\end{tabular}
\caption{\label{tbl:regional_cloudflare} Median mapping latencies for Cloudflare-CDN by region}
\end{table*}

\begin{table*}
\footnotesize
    \centering
\begin{tabular}{|c|c|c|c|c|c|c|c|c|}
\hline
\multirow{2}{*}{Region} &
\multicolumn{2}{|c|}{Google} & \multicolumn{2}{|c|}{Cloudflare-R} & \multicolumn{2}{|c|}{OpenDNS} & \multicolumn{2}{|c|}{Quad9} \\ \cline{2-9}
& IPv4 & IPv6 & IPv4 & IPv6 & IPv4 & IPv6 & IPv4 & IPv6 \\ \hline

Global (1717 Probes) & 11.40 & 12.10 & 11.34 & 11.70 & 11.37 & 11.65 & 13.90 & 13.39 \\ \hline 

Africa (32 Probes) & 19.37 & 20.87 & 20.04 & 18.16 & 20.04 & 18.15 & 49.27 & 18.20 \\ \hline 

Asia (212 Probes) & 12.80 & 18.34 & 12.25 & 16.00 & 13.16 & 15.97 & 14.89 & 22.57 \\ \hline 

Europe (1092 Probes) & 10.64 & 10.95 & 10.54 & 10.60 & 10.61 & 10.62 & 13.34 & 12.21 \\ \hline 

Oceania (65 Probes) & 15.48 & 15.22 & 15.48 & 15.19 & 15.47 & 14.64 & 15.58 & 15.13 \\ \hline 

N. America (271 Probes) & 12.89 & 13.41 & 12.78 & 13.16 & 12.95 & 13.38 & 12.66 & 13.50 \\ \hline 

S. America (45 Probes) & 29.33 & 29.38 & 35.28 & 31.47 & 29.49 & 29.05 & 110.16 & 107.12 \\ \hline 
\end{tabular}
\caption{\label{tbl:regional_edgecast}Median mapping latencies for Edgecast by region}
\end{table*}

\begin{table*}
    \centering
\footnotesize
\begin{tabular}{|c|c|c|c|c|c|c|c|c|}
\hline
\multirow{2}{*}{Region} &
\multicolumn{2}{|c|}{Google} & \multicolumn{2}{|c|}{Cloudflare-R} & \multicolumn{2}{|c|}{OpenDNS} & \multicolumn{2}{|c|}{Quad9} \\ \cline{2-9}
& IPv4 & IPv6 & IPv4 & IPv6 & IPv4 & IPv6 & IPv4 & IPv6 \\ \hline

Global (1717 Probes) & 10.01 & 10.97 & 9.91 & 10.91 & 10.02 & 10.80 & 15.48 & 17.53 \\ \hline 

Africa (32 Probes) & 7.72 & 18.13 & 5.69 & 18.15 & 7.75 & 18.14 & 7.76 & 16.34 \\ \hline 

Asia (212 Probes) & 9.31 & 12.51 & 9.82 & 12.13 & 9.88 & 11.94 & 70.45 & 84.24 \\ \hline 

Europe (1092 Probes) & 10.11 & 10.50 & 10.00 & 10.49 & 10.11 & 10.44 & 13.65 & 15.33 \\ \hline 

Oceania (65 Probes) & 5.52 & 8.75 & 5.75 & 8.70 & 5.57 & 8.97 & 21.10 & 21.38 \\ \hline 

N. America (271 Probes) & 10.51 & 12.03 & 10.49 & 11.92 & 10.33 & 11.84 & 15.31 & 18.09 \\ \hline 

S. America (45 Probes) & 13.85 & 15.89 & 11.39 & 15.21 & 13.61 & 16.56 & 108.65 & 108.80 \\ \hline 
\end{tabular}
\caption{\label{tbl:regional_fastly}Median mapping latencies for Fastly by region}
\end{table*}

\section{Up-to-date and Client-Specific Resolver Assessments}

Because the foregoing analysis is based 
on a one-time measurement campaign, it will
necessarily become obsolete as the platforms under study 
evolve. Further, although we have attempted to assemble a large and
diverse set of vantage points, our analysis provides 
{\em general trends} in resolvers performance, and the experience of
{\em a particular client} may deviate from these trends.
Consequently, we augment our study with two applications.  First, we
schedule monthly repeats of our measurement campaign and provide a web
portal \cite{nick-portal} that generates key results of our analysis
customized to user queries.  Second, we make available an
application \cite{Joey-app} that one can
download and run on their own computer, which compares the performance
of the four public resolvers to each other and to the ISP resolver
from the vantage point of their own computer.

\noindent
{\bf Recurrent measurements} Our instrumentation for recurrent
resolver analysis consists of two main components. One component implements a 
    workflow that automatically replicates the multi-step data collection
    described in this paper: it runs the script to identify websites
    accelerated by the (statically defined set of) CDNs under study (10 sites for Akamai and 5 sites for every other CDN) and
    selects one CDN-delivered URL per such site, runs the script to verify that the probes\footnote{Our current implementation only consider the 428 probes that were originally identified as suitable for our measurement.  Reconstructing this list from the entire set of RIPE Atlas probes to include newly deployed probes is future work.} still include ISP resolvers in their resolver lists, and then conducts the DNS
    and SSL measurements from these probes using the ISP and 
    (statically defined set of) public resolvers under study. The
    other component is a web portal a user can interact with to
    observe the results.  The portal allows the user to specify the
    time (month and year) for which they would like to see the
    results, the CDN whose mapping quality they would like to explore
    (or select all available CDNs), the protocols (IPv4, IPv6, or both) to be included
    in the analysis, and whether they would like the global view or
    focus on a particular region.   The portal then uses the data
    collected by the first component to generate custom CDF graphs and
    statistics for the user-specified parameters.  

 \noindent
    {\bf Client-specific measurements} To compare performance of
    different resolvers for the specific client environment, we have
    implemented a python application that the user can download and
    run on their own device.  By default, the app explores the
    same set of resolvers and CDNs as considered in the rest of our 
    study, using the latest set of CDN-delivered URLs produced by our
    recurrent measurement instrumentation to measure DNS and mapping
    latencies.  However, the user can augment both the resolver set
    and the CDNs by specifying additional resolver IP addresses and
    CDN-delivered URLs in configuration files.  Note that the user's
    ISP may block DNS traffic to unauthorized resolvers, which may
    include some or all public resolvers the user attempts to
    explore.  The app will exclude these resolvers from the comparison,
    which is appropriate: since the user cannot utilize these resolvers
    in their environment, the performance of these resolvers is
    irrelevant. 
    
\section{Conclusion}
\label{sec:concl}

This paper investigates two key performance aspects of the interplay between public DNS resolution services and content delivery networks -- the latency of DNS queries for resolving CDN-accelerated hostnames and the quality of the user-to-edge server mappings obtained through a given resolution service — as measured by latency.  In particular, we consider four popular public DNS resolvers — Google DNS, OpenDNS, Cloudflare resolver, and Quad9 — as they interact with four influential CDNs — Akamai, Cloudflare CDN, Edgecast, and Fastly.  We find that Google DNS and OpenDNS lag  behind the Cloudflare resolver and, for some CDNs, Quad9 in terms of DNS latency, and trace the cause to drastically lower cache hit rates in Google and OpenDNS. 

A the same time, Google and OpenDNS have largely closed
 the gap with ISP resolvers in the quality of CDNs'
 client-to-edge-server mappings as measured by latency, while
 the Cloudflare resolver still shows some penalty with Akamai, and Quad9 exhibits  a noticeable penalty with all CDNs in the study except for
the Cloudflare CDN, which does not use DNS to map clients to servers. 

Finally, we observe that the performance of public resolvers substantially depends on the region of the client.  For example, Google DNS has much lower DNS latency, for all four CDNs in the study, than OpenDNS in South America, but the reverse is true in Africa.  Similarly, while we do not observe IPv6 penalty in user mapping latencies globally, this penalty is present in Asia, as well as in Africa in the case of Fastly.  Furthermore, as observed on the example of Quad9 performance in South America, the performance of public resolvers may change qualitatively as their platforms evolve and new partnerships emerge. These observations highlight the need for ongoing reassessment of the performance of DNS resolution services and for comparison of these services from an individual client perspective.  Accordingly, we contribute a system for ongoing monthly measurements and a Web portal for flexibly querying the results, as well as an app a user can download to assess the resolvers performance from the user's specific vantage point.

\bibliographystyle{abbrv}
\bibliography{references,references_Nick}

\begin{thebibliography}{10}

\bibitem{IDC_MarketScape}
G.~Abdo and M.~Fremeijer.
\newblock {IDC} {MarketScape}: Worldwide commercial {CDN} 2019 vendor
  assessment.
\newblock https://www.speedykorea.com/pdf/idc\_report.pdf, 2019.

\bibitem{ager2010comparing}
B.~Ager, W.~M{\"u}hlbauer, G.~Smaragdakis, and S.~Uhlig.
\newblock {Comparing DNS Resolvers in the Wild}.
\newblock In {\em ACM IMC}, pages 15--21, 2010.

\bibitem{akamai-distribution}
Why {Akamai}.
\newblock https://www.akamai.com/why-akamai, 2023.

\bibitem{full_paper}
R.~Al-Dalky and M.~Rabinovich.
\newblock Revisiting comparative performance of {DNS} resolvers in the {IPv6
  and ECS} era.
\newblock {arXiv preprint arXiv:2007.00651}; available at
  http://arxiv.org/abs/2007.00651, 2020.

\bibitem{alzoubi2013performance}
H.~A. Alzoubi, M.~Rabinovich, and O.~Spatscheck.
\newblock {Performance Implications of Unilateral Enabling of IPv6}.
\newblock In {\em PAM}, pages 115--124. Springer, 2013.

\bibitem{bajpai2015}
V.~Bajpai, S.~J. Eravuchira, and J.~Sch{\"o}nw{\"a}lder.
\newblock Lessons learned from using the ripe atlas platform for measurement
  research.
\newblock {\em ACM SIGCOMM Computer Communication Review}, 45(3):35--42, 2015.

\bibitem{bajpai2016measuring}
V.~Bajpai and J.~Sch{\"o}nw{\"a}lder.
\newblock {Measuring the Effects of Happy Eyeballs}.
\newblock In {\em ACM Applied Networking Research Workshop}, pages 38--44,
  2016.

\bibitem{chen2015end}
F.~Chen, R.~K. Sitaraman, and M.~Torres.
\newblock {End-User Mapping: Next Generation Request Routing for Content
  Delivery}.
\newblock {\em ACM SIGCOMM Computer Comm. Review}, 45(4):167--181, 2015.

\bibitem{Cisco_VNI}
{Cisco Visual Networking Index: Forecast and Methodology, 2016–2021}.
\newblock
  https://www.cisco.com/c/en/us/solutions/colla\-teral/service-provider/visual-networking-index-vni/complete-white-paper-c11-481360.html\#\_Toc484813991,
  2017.

\bibitem{cisco-cdn-traffic}
{Cisco Visual Networking Index: Forecast and Trends, 2017–2022}.
\newblock
  https://twiki.cern.ch/twiki/pub/HEPIX/TechwatchNetwork/HtwNetworkDocuments/\\white-paper-c11-741490.pdf,
  2019.

\bibitem{cloudflare-locations}
The {Cloudflare} global network.
\newblock https://www.cloudflare.com/network/, 2023.

\bibitem{rfc7871}
C.~Contavalli, W.~van~der Gaast, D.~Lawrence, and W.~Kumari.
\newblock {Client Subnet in DNS Queries}.
\newblock RFC 7871, May 2016.

\bibitem{CAZ+14}
J.~Czyz, M.~Allman, J.~Zhang, S.~Iekel-Johnson, E.~Osterweil, and M.~Bailey.
\newblock {Measuring IPv6 Adoption}.
\newblock In {\em ACM SIGCOMM}, Aug. 2014.

\bibitem{google_cache}
M.~Daoudi.
\newblock {Performance: How Long Does a Second Actually Last?}
\newblock https://dzone.com/articles/performance-how-long-does, DZone
  Performance Zone, JUNE 2012.

\bibitem{DZone}
M.~Daoudi.
\newblock {A Complete Performance Breakdown of 1.1.1.1: Cloudflare's Public DNS
  Resolver}.
\newblock
  https://dzone.com/articles/a-complete-performance-breakdown-of-1111-cloudflar-1,
  DZone Performance Zone, APR 2018.

\bibitem{de2019passive}
W.~B. De~Vries, R.~van Rijswijk-Deij, P.-T. de~Boer, and A.~Pras.
\newblock Passive observations of a large {DNS} service: 2.5 years in the life
  of {Google}.
\newblock {\em IEEE Transactions on Network and Service Management},
  17(1):190--200, 2019.

\bibitem{fastly-anycast-performance}
Fastly documentation: Using {Fastly} with apex domains.
\newblock https://docs.fastly.com/en/guides/using-fastly-with-apex-domains,
  2023.

\bibitem{our_data}
\url{https://www.dropbox.com/sh/ioswha9fha3gmxn/AABVNt78urpVshVBbQvAGfvKa?dl=0}.

\bibitem{holterbach2015}
T.~Holterbach, C.~Pelsser, R.~Bush, and L.~Vanbever.
\newblock Quantifying interference between measurements on the {RIPE Atlas}
  platform.
\newblock In {\em ACM Internet Measurement Conference}, 2015.

\bibitem{hours2016study}
H.~Hours, E.~Biersack, P.~Loiseau, A.~Finamore, and M.~Mellia.
\newblock {A study of the impact of DNS resolvers on CDN performance using a
  causal approach}.
\newblock {\em Computer Networks}, 109:200--210, 2016.

\bibitem{huang2011public}
C.~Huang, D.~A. Maltz, J.~Li, and A.~Greenberg.
\newblock {Public DNS system and Global Traffic Management}.
\newblock In {\em INFOCOM}, pages 2615--2623, 2011.

\bibitem{nick_thesis}
N.~Kernan.
\newblock A large-scale assessment of dns resolution services.
\newblock Master's thesis, Case Western Reserve University, 2023.

\bibitem{special-hostname}
M.~Korf.
\newblock Introducing new {WhoAmI} tool {DNS} resolver information.
\newblock
  https://www.akamai.com/blog/developers/introducing-new-whoami-tool-dns-resolver-information,
  2018.

\bibitem{law2008empirical}
Y.-N. Law, M.-C. Lai, W.~L. Tan, and W.~C. Lau.
\newblock Empirical performance of {IPv6 vs. IPv4} under a dual-stack
  environment.
\newblock In {\em 2008 IEEE International Conference on Communications}, pages
  5924--5929. IEEE, 2008.

\bibitem{DNS_resolvers_NANOG18}
A.~Medina.
\newblock {Comparing the Performance of Public DNS Resolvers}.
\newblock {Presentation at NANOG-74 Meeting; available at
  youtube.com/watch?v=gSV5jsZiYiQ}, Sept. 2018.

\bibitem{ttl-violations}
G.~Moura.
\newblock {DNS TTL} violations in the wild - measured with {RIPE Atlas}.
\newblock
  https://labs.ripe.net/author/giovane\_moura/dns-ttl-violations-in-the-wild-measured-with-ripe-atlas/,
  2017.

\bibitem{nikkhah2011assessing}
M.~Nikkhah, R.~Gu{\'e}rin, Y.~Lee, and R.~Woundy.
\newblock Assessing {IPv6} through {Web} access a measurement study and its
  findings.
\newblock In {\em Proceedings of the Seventh conference on emerging Networking
  EXperiments and Technologies}, pages 1--12, 2011.

\bibitem{otto2012content}
J.~S. Otto, M.~A. S{\'a}nchez, J.~P. Rula, and F.~E. Bustamante.
\newblock {Content Delivery and the Natural Evolution of DNS: Remote DNS
  Trends, Performance Issues and Alternative Solutions}.
\newblock In {\em Proceedings of the 2012 Internet Measurement Conference},
  pages 523--536. ACM, 2012.

\bibitem{cloudflare-anycast2}
M.~Prince.
\newblock Load balancing without load balancers.
\newblock
  https://blog.cloudflare.com/cloudflares-architecture-eliminating-single-p/,
  2013.
\newblock Accessed on 07/10/2024.

\bibitem{quad9-south-america}
Quad9 selects {EdgeUno} to expand their services throughout {Latin America}.
\newblock
  https://www.quad9.net/news/press/quad9-selects-edgeuno-latin-america/, 2022.

\bibitem{quad9-locations}
Quad9: Locations.
\newblock https://quad9.net/service/locations/, 2023.

\bibitem{randall2020trufflehunter}
A.~Randall, E.~Liu, G.~Akiwate, R.~Padmanabhan, G.~M. Voelker, S.~Savage, and
  A.~Schulman.
\newblock Trufflehunter: Cache snooping rare domains at large public {DNS}
  resolvers.
\newblock In {\em ACM Internet Measurement Conference}, pages 50--64, 2020.

\bibitem{RIPE_Atlas}
{Welcome to RIPE Atlas}.
\newblock https://atlas.ripe.net/.

\bibitem{sanchez2013dasu}
M.~A. S{\'a}nchez, J.~S. Otto, Z.~S. Bischof, D.~R. Choffnes, F.~E. Bustamante,
  B.~Krishnamurthy, and W.~Willinger.
\newblock {Dasu: Pushing Experiments to the Internet's Edge}.
\newblock In {\em NSDI}, pages 487--499, 2013.

\bibitem{rfc8305}
D.~Schinazi and T.~Pauly.
\newblock {Happy Eyeballs Version 2: Better Connectivity Using Concurrency}.
\newblock RFC 8305, RFC Editor, December 2017.

\bibitem{Schomp2020AkamaiDNS}
K.~Schomp, O.~Bhardwaj, E.~Kurdoglu, M.~Muhaimen, and R.~K. Sitaraman.
\newblock Akamai {DNS}: Providing authoritative answers to the world's queries.
\newblock In {\em ACM SIGCOMM}, page 465–478, 2020.

\bibitem{schomp2013measuring}
K.~Schomp, T.~Callahan, M.~Rabinovich, and M.~Allman.
\newblock {On Measuring the Client-Side DNS Infrastructure}.
\newblock In {\em IMC}, pages 77--90, 2013.

\bibitem{cymru}
{Team Cymru}.
\newblock {IP To ASN Mapping}.
\newblock \url{http://www.team-cymru.com/IP-ASN-mapping.html}.

\bibitem{sipat_thesis}
S.~Triukose.
\newblock {\em A PEER-TO-PEER INTERNET MEASUREMENT PLATFORM AND ITS
  APPLICATIONS IN CONTENT DELIVERY NETWORKS}.
\newblock PhD thesis, Case Western Reserve University, 2014.

\bibitem{wang2005understanding}
Y.~Wang, S.~Ye, and X.~Li.
\newblock Understanding current {IPv6} performance: a measurement study.
\newblock In {\em 10th IEEE Symposium on Computers and Communications
  (ISCC'05)}, pages 71--76. IEEE, 2005.

\bibitem{Joey-app}
{LDNS Explorer}.
\newblock https://pypi.org/project/ldns-explorer/.

\bibitem{nick-portal}
{Web Portal for Performance Measurements of DNS Resolver Platforms }.
\newblock https://dns-web-portal.netlify.app/.

\bibitem{zhou2008ipv6}
X.~Zhou, M.~Jacobsson, H.~Uijterwaal, and P.~Van~Mieghem.
\newblock {IPv6} delay and loss performance evolution.
\newblock {\em International Journal of Communication Systems}, 21(6):643--663,
  2008.

\end{thebibliography}

\end{document}